\documentclass[10pt,journal,compsoc]{IEEEtran}

\include{0header}

\usepackage{upgreek}

\newcommand{\StateSpace}{\mathcal{S}}
\newcommand{\state}{s}
\newcommand{\allstates}{S}
\newcommand{\statetransition}{\texttt{statenext}}

\newcommand{\epoch}{t}
\newcommand{\nepochs}{T}
\newcommand{\EpochSet}{\mathcal{T}}

\newcommand{\ninnerepochs}{O}
\newcommand{\timeinterval}{\Delta}
\newcommand{\fixingtime}{c}

\newcommand{\bug}{b}
\newcommand{\BugSet}{\mathcal{B}}
\newcommand{\LDA}{\texttt{LDA}}

\newcommand{\Due}{\texttt{due}}

\newcommand{\developer}{d}
\newcommand{\DeveloperSet}{\mathcal{D}}

\newcommand{\Sch}{\texttt{sch}}
\newcommand{\Exp}{\texttt{exp}}
\newcommand{\variable}{y}

\newcommand{\valuefunction}{V}
\newcommand{\nduedate}{U}

\newcommand{\DuedateSet}{\mathcal{U}}
\newcommand{\postponed}{p}
\newcommand{\ActionSet}{\mathcal{A}}
\newcommand{\action}{a}
\newcommand{\allactions}{A}
\newcommand{\coefficient}{f}

\newcommand{\slack}{h}
\newcommand{\iteration}{n}
\newcommand{\niteration}{N}
\newcommand{\innerteststep}{M}
\newcommand{\exoginfo}{\xi}
\newcommand{\policy}{\pi}
\newcommand{\PolicySet}{\Pi}
\newcommand{\discount}{\gamma}
\newcommand{\Reward}{R}
\newcommand{\stepsize}{\alpha}
\newcommand{\lambdahram}{\eta}
\newcommand{\BAKFestimate}{\theta}
\newcommand{\BAKFappbias}{\beta}
\newcommand{\BAKFsqrpred}{\delta}
\newcommand{\BAKFerrorstep}{\nu}
\newcommand{\BAKFnoise}{\sigma}
\newcommand{\BAKFvariance}{\lambda}

\begin{document}\sloppy

\title{ADPTriage: Approximate Dynamic Programming for Bug Triage}




\author{Hadi~Jahanshahi, 
        Mucahit~Cevik, 
        Kianoush~Mousavi, 
        and~Ay\c{s}e~Ba\c{s}ar,
\IEEEcompsocitemizethanks{\IEEEcompsocthanksitem H. Jahanshahi, M. Cevik, and A. Ba\c{s}ar are with the Data Science Lab, Toronto Metropolitan University (Formerly Ryerson University), Toronto, Ontario, M5B 2K3, Canada.\protect\\
E-mail: hadi.jahanshahi@ryerson.ca
\IEEEcompsocthanksitem K. Mousavi is with Industrial Engineering department at the University of Toronto.}
}

\markboth{}
{Jahanshahi \MakeLowercase{\textit{et al.}}: ADPTriage: Approximate Dynamic Programming for Bug Triage}


\IEEEtitleabstractindextext{%
\begin{abstract}
Bug triaging is a critical task in any software development project. It entails triagers going over a list of open bugs, deciding whether each is required to be addressed, and, if so, which developer should fix it. However, the manual bug assignment in issue tracking systems (ITS) offers only a limited solution and might easily fail when triagers must handle a large number of bug reports. 
During the automated assignment, there are multiple sources of uncertainties in the ITS, which should be addressed meticulously.
In this study, we develop a Markov decision process (MDP) model for an online bug triage task. 
In addition to an optimization-based myopic technique, we provide an ADP-based bug triage solution, called ADPTriage,
which has the ability to reflect the downstream uncertainty in the bug arrivals and developers' timetables. 
Specifically, without placing any limits on the underlying stochastic process, this technique enables real-time decision-making on bug assignments while taking into consideration developers' expertise, bug type, and bug fixing time. 
Our result shows a significant improvement over the myopic approach in terms of assignment accuracy and fixing time. 
We also demonstrate the empirical convergence of the model and conduct sensitivity analysis with various model parameters. 
Accordingly, this work constitutes a significant step forward in addressing the uncertainty in bug triage solutions. 
\end{abstract}

}

\maketitle

\IEEEdisplaynontitleabstractindextext

\IEEEpeerreviewmaketitle

\IEEEraisesectionheading{\section{Introduction}\label{sec:introduction}}
\IEEEPARstart{B}{ug} repositories and issue tracking systems (ITS) are commonly used to track and address software issue reports, whether they are feature improvement requests or bugs arising during the testing or maintenance phase. To manage these requests, open-source software projects mainly rely on issue-tracking platforms, such as Bugzilla, Jira, and GitHub~\citep{Aung2022}. Bug triage task involves promptly prioritizing bugs and assigning them to appropriate developers. It is preceded by examining the validity of the bugs, determining the possible missing information in the bug report, checking possible duplicate bug reports, setting the severity of bugs, and assigning them to a proper developer~\citep{Banerjee2017, Aung2022}. 
As such, bug triage is deemed as a challenging task that directly affects software quality. 
Since numerous bugs are reported to open-source software systems every day, the manual triage task in such projects is prone to subjective or erroneous decisions. 
Accordingly, our focus in this study is on the bug assignment task, which involves assigning bugs to the most appropriate developers at the right time. 

Researchers proposed many bug triage approaches to overcome the issues of manually assigning bugs to developers. As bug triage is a multifaceted problem involving different sub-tasks, each study concentrates on a particular aspect of the problem. Most of the literature is dedicated to improving the accuracy of the bug assignment by proposing appropriate developers based on the textual information of the bugs~\citep{anvik2006should, lee2017applying, Xi2018, mani2019deeptriage, jahanshahi2020Wayback, Zaidi2022}. On the other hand, some studies consider various other dimensions of the problem. For instance, ~\citet{park2011costriage} developed the CosTriage algorithm that, besides the accuracy of the assignment (i.e., assigning to the right person), considers the fixing cost of the bugs. CosTriage employs the content-based recommendation and collaborative filtering recommender to construct developers' profiles and approximate the fixing time of each bug type. \citet{kashiwa2020} enhanced CosTriage by proposing an integer programming-based solution that incorporates the developers' suitability, bug fixing time, and the software release dates. Although \citet{kashiwa2020}'s approach covered several important aspects of the bug triage, \citet{jahanshahi2022sdabt} further extended the previous methods by incorporating the developers' schedules and the bug dependencies. 
Similarly, \citet{Yadav2022} highlighted the importance of fair bug distribution among developers. Their method manages to avoid assigning bugs to developers who are overloaded by taking into account the current workload and expertise of developers.

Although previous works proposed different algorithms to accommodate various aspects of the bug triage, to the best of our knowledge, the \textit{uncertainty} in the ITS is not yet explored in any model. For instance, the number of bugs reported to the system does not follow a particular pattern to be easily predicted~\citep{Wang2012}. Thus, we are not certain whether a severe bug will be reported in the upcoming day so that triagers/developers can plan ahead instead of being involved in less important open bugs in the system. On the other hand, the developers' schedules are another source of uncertainty, which may constantly get updated in open-source software systems since not all developers are as dedicated as the ones in a proprietary software system. Therefore, the developer may be available the next day or not, depending on their schedule/availability. Accordingly, a triage model for the ITS should account for such uncertainties. 

To the best of our knowledge, our study constitutes the first work that provides a stochastic model for the bug triage problem to maximize the long-run returns. 
The return is defined based on matching developers' expertise and bug types and the corresponding action of either assigning the open bugs at the current timestamp or postponing them with the expectation of a better match for the upcoming bugs. We introduce a novel Markov Decision Process (MDP) model for the bug triage problem and leverage Approximate Dynamic Programming (ADP) to capture the uncertainty in the bug triage environment. Our proposed method, which we refer to as \texttt{ADPTriage}, not only assigns the bugs to the most appropriate developers or postpones them to the future but also determines the assignment timing according to the likelihood of having a particular bug type in the system and possible changes in developers' schedules in the future timestamps. 

The rest of the paper is organized as follows. The background of the main approaches employed in our proposed model is briefly discussed in Section~\ref{sec:background3}. The \texttt{ADPTriage} technique and bug report datasets utilized in our research are described in Section~\ref{sec:methodology}. The numerical experiments and comparisons with the baseline methods are presented in Section~\ref{sec:results3}.
Section~\ref{sec:threats3} discusses the study limitations, followed by the literature review in Section~\ref{sec:related_works3}. Finally, Section~\ref{sec:conclusion3} summarizes the paper and elaborates on potential future research directions.

\section{Background} \label{sec:background3}
In this section, we discuss the background of the methodological approaches utilized in our proposed model.

\subsection{LDA}\label{sec:LDA}
The Latent Dirichlet allocation (LDA) is an unsupervised, probabilistic topic modeling technique that uses word clusters and frequencies to identify topics in corpora~\citep{blei2003LDA}. LDA presumes that the document has $n$ many subjects, to one of which each word is assigned. LDA is commonly used in bug triage problems to determine the bug type given the bugs' textual information, where the former relates to subjects and the latter to documents. Hence, after removing the stop words, we extract bug descriptions and summaries from a bug report (as textual information) and construct a bag of words. For consistency, we follow the same steps as previous studies~\citep{park2011costriage, kashiwa2020}. In this paper, by referring to the ``bug type'', we mean their LDA category. We also estimate the experience of each developer for each LDA category. We utilize Arun's technique to determine the best number of LDA categories~\citep{arun2010}, and then, the average bug fixing time of each developer given each category is computed. Finally, we employ a collaborative filtering recommender to approximate the missing values. For the sake of consistency with previous studies, we follow the exact steps introduced by~\citet{park2011costriage}.

\subsection{Approximate Dynamic Programming} \label{sec:ADP}

\begin{figure}[!ht]
    \centering
    \includegraphics[width=0.25\textwidth]{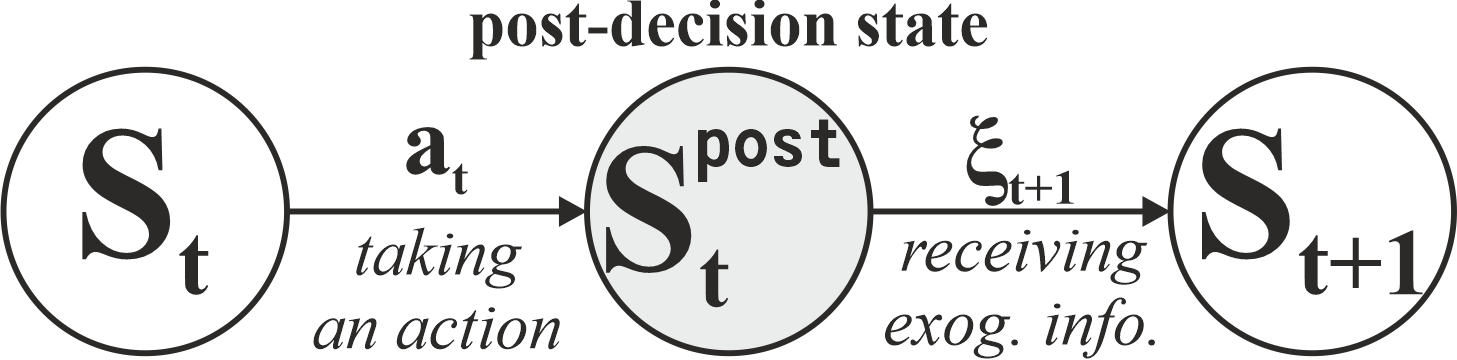}
    \caption{A typical state transition in MDP models}
    \label{fig: transition}
\end{figure}

The general task of optimizing decisions over a time horizon extends to varied backgrounds and domains. Said problems are known as \textit{sequential decision-making problems}, and given sets of states, decisions, and time, may be formulated through the use of a Markov Decision Process (MDP)~\citep{powell2022}. Such a model is commonly utilized in resource allocation problems, e.g., allocating fleets of vehicles~\citep{Shah2020}, manpower planning problem~\citep{bradshaw2016}, and assigning a task to an expert~\citep{Erdelyi2010}. In a typical MDP problems , an agent, according to its current state, takes an action. That action moves the agent to a new state, called the post-decision state, where the agent receives new information from the environment, also called exogenous information, and accordingly, it proceeds to the next state (see Figure~\ref{fig: transition}). 

Generally, MDP models consist of seven main components. The first two, the state and action space, $\state_\epoch \in \StateSpace_\epoch$ and $\action_\epoch \in \ActionSet_\epoch$, respectively, help to define the state of the system as well as the set of feasible decisions (or actions) at each time step $\epoch$. The next component, exogenous information, $\exoginfo_{\epoch+1}$, independently arrives in the system and is used along with the two aforementioned components as input for moving the state of the system forward in time. Such forward movement is explicitly characterized through the use of a transition function, $\state_{\epoch+1} = \statetransition(\state_\epoch,\action_\epoch,\exoginfo_{\epoch+1})$, along with transition probabilities $\mathbb{P}(\state_{\epoch+1}|\state_\epoch,\action_\epoch)$, which probabilistically shift the system state based on its current state at time $\epoch$, the corresponding action(s) taken, and the exogenous information. The objective function looks to optimize the expected discounted reward/penalty and provides a policy. More specifically, a policy, $\policy \in \PolicySet$ (where $\PolicySet$ is the set of all decision policies), is a decision function $\allactions^{\policy}_\epoch(\state_\epoch)$ which tells an agent which feasible action $\action_\epoch$ to take at every time step $\epoch$ in each given state $\state_\epoch$. The ultimate goal is to find a policy that optimizes the total expected reward/cost received over a given time horizon. It can be written concisely as
\begin{align}
\max_{\policy \in \PolicySet}\mathbb{E}^{\policy} \left\{ \sum_{\epoch=1}^{\nepochs} \discount_\epoch \Reward_\epoch\big( \allstates_\epoch, \allactions_{\epoch}^{\policy}(\allstates_\epoch) \big) \right\}, \forall \exoginfo_\epoch \in \Xi  \label{eq:expected_reward}
\end{align}
in which $\discount$ represents the discounting factor, $\Reward_\epoch$ is the expected reward given being in state $\state_\epoch$ and taking action $\action_\epoch$. We discretize the time horizon into $\nepochs$ epochs.

Backward Dynamic Programming (DP) is one of the possible options for solving such discrete stochastic processes. Its solution methodology seeks to break down the overarching MDP problem into sub-problems, solving each sub-problem through the use of recursive equations which attempt to capture both the immediate and downstream value of being in a given state at a set time. The optimal policy for such sub-problems then provides the optimal solution to the mentioned MDP~\citep{sutton2018reinforcement}. Such a formulation may be written using Bellman’s optimality equation as follows:
\begin{align}
\valuefunction_{\epoch}(\state_\epoch) = \max_{\action_\epoch \in \ActionSet_\epoch} \big(\Reward_\epoch(\state_\epoch,\action_\epoch) + \discount \mathbb{E}[\valuefunction_{\epoch+1}(\state_{\epoch+1})|\state_\epoch]\big),
\end{align}
where $\valuefunction_{\epoch}(\state_\epoch)$ is the value function and $\state_{\epoch+1} = \statetransition(\state_\epoch,\action_\epoch,\exoginfo_{\epoch+1})$. Here, at each timestep, we define the value of a given state through the action that optimizes its immediate and expected discounted downstream reward. However, finding the exact solution of such an equation (typically through the use of backward dynamic programming) proves to be difficult and/or intractable for large problems due to the three ``curses of dimensionality''~\citep{powell2007approximate}. More specifically, they are related to the presence of (i) a large multidimensional state space making value function approximation difficult, (ii) a large multidimensional action space hindering optimal decision making, and (iii) a multidimensional outcome space impeding computing the expectation of \textit{future} rewards.

ADP, which is a powerful algorithmic framework used to solve such large-scale discrete-time MDP problems, helps to address the aforementioned three curses of dimensionality. To address (i), the concept of a post-decision state, ${\state^{\texttt{post}}_\epoch}$, is introduced that helps to define the state of the system after an action has been taken at time $\epoch$ but prior to the arrival of any exogenous information at the beginning of the next time-step $\epoch+1$. This allows us to break the transition function into two components, namely, $\state^{\texttt{post}}_\epoch=\texttt{statepost}(\state_\epoch,\action)$ which defines the transition to the post-decision state, and $\state_{\epoch+1}=\texttt{statenext}(\state^{\texttt{post}}_\epoch,\exoginfo_{\epoch+1})$ which characterizes the transition into the next state at time $\epoch+1$ after the arrival of exogenous information. To address (ii),  we estimate the value function of the post-decision state with the linear approximation of individual bug and developer types. It sums over the expected future values of postponed bugs and the expected values of the available developer at epoch $\epoch+1$.

To address (iii), sample paths are generated over the planning horizon, and a forward dynamic programming approach is taken to solve the Bellman equations, stepping forward in time and repeating the process for multiple iterations. Thus, using the post-decision states, we are able to break the Bellman equation into two parts:
\begin{align}
& \valuefunction(\state_\epoch) = \max_{\action_\epoch \in \allactions_\epoch} \big(\Reward_\epoch(\state_\epoch,\action_\epoch) + \discount \valuefunction^\texttt{statepost}(\state_\epoch^\texttt{statepost})\big) \label{eq:ValueOfPostdecisionState}\\
& \valuefunction^\texttt{statepost}(\state_\epoch^\texttt{statepost}) = \mathbb{E} \big[ \valuefunction(\state_{\epoch+1})| \state_\epoch^\texttt{statepost},\exoginfo_{\epoch+1} \big]. \label{eq:ValueOfNextState}
\end{align}

Equation~\eqref{eq:ValueOfPostdecisionState} becomes deterministic, making it easier to solve, and Equation~\eqref{eq:ValueOfNextState} is approximated and updated by stepping forward in time and observing sample realizations of exogenous information. It will be explained in more detail in Section~\ref{sec:adpformulation}.

\section{Problem Formulation}\label{sec:methodology}
We present an online bug triage system that assigns bugs to active developers in issue-tracking systems. 
In particular, we focus on open-source software projects, e.g., {\scshape{EclipseJDT}}, {\scshape{GCC}}, and {\scshape{Mozilla}}.
In open-source software systems, bug arrivals and developers' availabilities demonstrate dynamic behaviour which may evolve over time. We assume that bugs can be clustered into different subcategories according to their textual title and description using the LDA. These bugs arrive at the system based on a stochastic process, with an expected assignment deadline that may vary according to their priority/severity. 
It is desirable to triage a bug before its deadline. 
Hence, a late assignment cost is associated with those bugs remaining in the system longer than their due date. 
Assuming that we have a predefined project horizon, we discretize it into time intervals. 
We denote the discrete set of decision-making epochs by $\EpochSet \coloneqq \{1, 2, \dots, \nepochs\}$. Decisions are taken at the start of each time interval $\epoch$ while exogenous information is observed between two intervals. We assume that the time intervals between epochs, denoted by $\timeinterval$, are equal. Without losing the generalizability, equispaced decision intervals may approach 0, making the model an online bug recommendation system. Similar to \citet{kashiwa2020} and \citet{jahanshahi2022sdabt}, $\timeinterval$ in our experiments is one day. Therefore, the model assigns bugs to the proper developers once a day. The granularity of this epoch length can be easily adjusted without violating the model generalizability. 

\subsection{Assumptions}
We make the following assumptions to construct an ADP solution for the bug triage problem.

\begin{itemize}\setItemSep{0.5em}
    \item We only evaluate \textit{active} developers and omit inactive developers because we do not have enough information about them~\citep{anvik2006should, park2016cost, kashiwa2020}. Less active ones may visit the ITS infrequently, and little may be known about their schedule and availability distribution. We define active developers as those whose bug fix number is higher than the interquartile range (IQR) of all developers' bug fix numbers, using IQR as a measure of central distribution~\citep{park2016cost}. We acknowledge that in the agile software industry, the list of active developers must undergo some updates regularly as some may leave the company. The number of active developers in {\scshape{EclipseJDT}}, {\scshape{GCC}}, and {\scshape{Mozilla}} is 16, 47, and 128, respectively.
    
    \item A Bug of type $\bug$ has the fixing time of $\fixingtime_\bug^\developer$ if fixed by a developer with the experience $\developer_\texttt{exp}$. Similar to the previous studies by \citet{kashiwa2020} and \citet{park2011costriage}, we utilize LDA for topic modeling and then find the average fixing time of each developer given the category. 
    For instance, we have 6, 5, and 5 bug types defined by Arun's technique for {\scshape{EclipseJDT}}, {\scshape{GCC}}, and {\scshape{Mozilla}}, respectively. 
    
    \item No more than one developer can fix a bug simultaneously.
    
    \item If developer $\developer$ agrees to fix bug $\bug$, they will be unavailable for the next $\fixingtime_\bug^\developer$ epochs. During those times, no new bugs can be allocated to them.
    
    \item Each developer determines their vacations or off-days beforehand so that their schedules become updated accordingly.
\end{itemize}

\subsection{Dataset}

We consider three large OSS projects in our experiments. There are enough bug reports for these well-established projects, {\scshape{EclipseJDT}}, {\scshape{GCC}} and {\scshape{Mozilla}}.
We collect the bug report data from the bug repositories using the Bugzilla REST API\footnote{\url{https://wiki.mozilla.org/Bugzilla:REST_API}}. 
It incorporates both general bug attributes and bug metadata change history. 
Table~\ref{tab:dataset3} shows the information of the extracted datasets. 
We use bug reports between 2010 to 2018 as the training set and between 2018 and 2020 as the testing set.

\begin{table*}[!ht]
\centering
\caption{{Summary information for the bug datasets. The training phase is between Jan. 1st, 2010 and Dec. 31st, 2017, while the testing phase includes the data from Jan. 1st, 2018 to Dec. 31st, 2019.}\label{tab:dataset3}}%
\resizebox{.85\linewidth}{!}{
    \begin{tabular}{lrr|rr|rr}
        \toprule
        & \multicolumn{2}{c|}{{\scshape{{EclipseJDT}}}} &
        \multicolumn{2}{c|}{{\scshape{{GCC}}}}  &
        \multicolumn{2}{c}{{\scshape{{Mozilla}}}} \\ 
         & \textbf{Training} & \textbf{Testing} &  \textbf{Training} & \textbf{Testing}  &  \textbf{Training} & \textbf{Testing}  \\
        \midrule
        \textbf{Total bugs reported} & 12,598 & 3,518 & 34,635 & 9,998 & 90,178 & 22,353\\
        \textbf{Total bug dependencies found} & 2,169 & 970 & 4,462 & 3,268 & 71,549 & 19,223 \\
        \textbf{Total relevant changes in the bugs' history} & 55,109 & 15,505 & 138,580 & 42,117 & 410,010 & 114,778\\
        \textbf{Mean and Median fixing time (days)} & (41.2, 3) & (15.7, 1) & (42.3, 3) & (42.0, 3) & (27.2, 5) & (12.6,4) \\
        \textbf{Minimum and Maximum fixing time (days)} & (1, 1,753) & (1, 423) & (1, 2,396) & (1, 681) & (1, 2,172) & (1, 550) \\ 
        \midrule
        \textbf{After cleaning} &  &  &  &  & & \\
            \qquad \textbf{0. Bugs that are not META bugs} & 12,598 & 3,518 & 34,634 & 9,996 & 89,354 & 21,912 \\
            \qquad \textbf{1. Bugs with resolved status} & 11,296 & 2,619 & 29,057 & 7,195 & 79,108 & 18,720 \\
            \qquad \textbf{2. Bugs assigned to active developers} & 3,795 & 1,491 & 27,088 & 6,719 & 10,948 & 6,365 \\
            \qquad \textbf{3. Bugs with known assignment date} & 3,021 & 1,348 & 8,628 & 3,072 & 6,768 & 3,947 \\
            \qquad \textbf{4. Bugs with acceptable fixing time} & 2,372 & 1,201 & 7,166 & 2,459 & 5,618 & 3,547 \\
        \bottomrule
    \end{tabular}
}
\end{table*}

We only take into account bugs that match the following criteria, similar to earlier research by ~\citet{kashiwa2020} and \citet{park2011costriage}. First, META bugs are removed as they are used only to cluster similar bugs and do not have their own test cases. Second, we only consider the fixed bugs. Third, the bugs fixed by less active developers are excluded. Fourth, if the assignment date of a bug is unknown, the bug is omitted. Lastly, bugs with a fixing time greater than Q3 + (1.5 $\times$ IQR) are considered outliers. Q3 is the third quartile of bug fixing times, and IQR is the interquartile range of the fixing times. The acceptable fixing time for {\scshape{EclipseJDT}}, {\scshape{GCC}}, and {\scshape{Mozilla}} is 21, 38.5, and 6 days, respectively. Moreover, regarding preprocessing the textual information (i.e., bug titles and descriptions), we use lemmatization, stop words, numbers and punctuation removal, and lengthy word elimination (i.e., longer than 20 characters) similar to previous studies~\citep{kashiwa2020,liu2016multi,park2016cost}.

\subsection{ADPTriage}

We next provide the components of our MDP model and details of the ADP approach.

\subsubsection{State variables}

The state of the system is defined by open bugs and available developers at each epoch as follows:

\begin{itemize}
    \item A developer is characterized by their state vector $\developer = \big(\developer_\Exp, \developer_\Sch \big)$, representing the experience of a developer in solving specific types of bugs according to their LDA category and the schedule of the developer. The experience is estimated based on their history of fixing bugs and the LDA algorithm (see Section~\ref{sec:LDA}). The schedule of a developer ($\Sch$) shows the number of epochs until the developer's availability. For instance, if it is 5, it means that after five epochs, the developer becomes available. In the same way, $\Sch=0$ shows that the developer is available in the current epoch.  $\DeveloperSet_\epoch$ is the set of all developers with their associated attributes at epoch $\epoch$.
    
    \item A two-dimensional attribute vector, $\bug = \big(\bug_\LDA, \bug_\Due \big)$, describes an open bug in the system, capturing its LDA category and deadline. LDA category is used to estimate the fixing time $\fixingtime_\bug^\developer$ of bug $\bug$ if assigned to developer $\developer$. On the other hand, the deadline attribute corresponds to the acceptable number of epochs left to assign the bug to a developer for its on-time assignment. The deadline attribute belongs to the set $\DuedateSet \coloneqq \{0, 1, \dots, \nduedate\}$. Therefore, each bug has at most $\nduedate$ epochs to be assigned (i.e., the maximum time of $\nduedate\times \timeinterval$ for its assignment since being reported to the ITS).
    Exceeding the due date will incur a relative cost to the system. We assume that if a bug is reported to the system between decision epochs $\epoch-1$ and $\epoch$, it enters the ITS at the beginning of epoch $t$ with
    \begin{align}
        \bug_\Due = \min ({\nepochs-\epoch-1, \nduedate 
        })
    \end{align}
    
    In other words, $\bug_\Due$ is initialized by the number of epochs left until the end of the project horizon and the maximum assignment time of a bug
    . At the end of each epoch, we reduce $\bug_\Due$ by 1 since it should always show the remaining time until the estimated due date. Therefore, if we pass the due date, $\bug_\Due$ will become negative. $\BugSet_\epoch$ is the set of all open bugs at epoch $\epoch$.
\end{itemize}

Let $\StateSpace_{\epoch\bug}^\texttt{bug}$ and $\StateSpace_{\epoch\developer}^\texttt{dev}$ be the number of open bugs with attribute vector $\bug$ at time $\epoch$ and the number of available developers with attribute $\developer$ at time $\epoch$, respectively. Then, the system state at $\epoch \in \EpochSet$ is defined as $\state_\epoch \coloneqq \big(\StateSpace_\epoch^\texttt{bug}, \StateSpace_\epoch^\texttt{dev}\big)$, incorporating state space of the open bugs $\StateSpace_\epoch^\texttt{bug} = (\StateSpace_{\epoch\bug}^\texttt{bug})_{\bug \in \BugSet}$ and state space of the available developers $\StateSpace_\epoch^\texttt{dev} = ( \StateSpace_{\epoch\developer}^\texttt{dev})_{\developer \in \DeveloperSet}$.

\subsubsection{Decision Variables}
Given the present system state $\state_\epoch$, we have two different types of decisions at each decision epoch $\epoch$. The first one consists of bugs deferred to future decision epochs in the hopes of being assigned more cost-effectively (for example, allocating to a developer that becomes available in $\epoch+1$ and is more expert in fixing this bug type). We define the variable $\postponed_{\epoch\bug}$ for bugs, denoting the number of bugs with attribute $\bug$ postponed to the next decision epoch. The second possible decision set would be the assignment to the available developers, consisting of the bugs to be fixed. Thus, $\variable_{\epoch\developer\bug}$ is the number of developers with attribute $\developer$ assigned to bugs with attribute $\bug$ at epoch $\epoch$.

With the above definition, we denote the decision tuple $\action_\epoch = \big(\variable_\epoch, \postponed_\epoch\big)$ at epoch $\epoch$, where $\variable_\epoch$ and $\postponed_\epoch$ are decision vectors of variables $\variable_{\epoch\developer\bug}$ and $\postponed_{\epoch\bug}$, respectively. We denote the feasible decision set $\ActionSet_\epoch(\state_\epoch)$ given the current state of the system $\state_\epoch \coloneqq \big(\StateSpace_\epoch^\texttt{bug}, \StateSpace_\epoch^\texttt{dev}\big)$. The tuple $\action^\epoch \in \ActionSet_\epoch(\state_\epoch)$ should satisfy the following constraints:
\begin{align}
    \sum_{\bug \in \BugSet_\epoch}{\variable_{\epoch\developer\bug}} + \slack_{\epoch\developer}  &= \StateSpace_{\epoch\developer}^\texttt{dev} &\forall \developer \in \DeveloperSet_{\epoch} \label{eq:assigned_devs}\\[0.1em]
    \sum_{\developer \in \DeveloperSet_{\epoch}}{\variable_{\epoch\developer\bug}} + \postponed_{\epoch\bug} & = \StateSpace_{\epoch\bug}^\texttt{bug} & \forall \bug \in \BugSet_\epoch \label{eq:assigned_bugs}, 
\end{align}
where $\slack_{\epoch\developer}$ is the slack of constraints~\eqref{eq:assigned_devs}. Constraints~\eqref{eq:assigned_devs} ensure that the number of developers with attribute $\developer$ assigned to all open bugs does not exceed the total number of available developers with that attribute type. Constraints~\eqref{eq:assigned_bugs} control the flow of the bugs, making sure that each bug with attribute $\bug$ is either assigned to a developer or postponed to the next timestamp. Constraints~\eqref{eq:assigned_devs} and \eqref{eq:assigned_bugs} have a totally unimodular constraint matrix, implying that integral solutions can be found at all extreme points of the obtained feasible region, as such the integrality restrictions on the $\variable_{\epoch\developer\bug}$ and $\postponed_{\epoch\bug}$ decisions can be relaxed~\citep{wolsey1999integer}.

\subsubsection{Cost Function}

The total cost incurred as a result of actions $\action_\epoch$ for the state $\state_\epoch$ at epoch $\epoch \in \EpochSet$ is calculated as
\begin{align}
    \text{Cost}(\state_\epoch, \action_\epoch) = \sum_{\bug \in \BugSet_\epoch}{\coefficient(\bug_\texttt{due}, \epoch)\postponed_{\epoch\bug}} +
    \sum_{\developer \in \DeveloperSet_{\epoch}} {\sum_{\bug \in \BugSet_\epoch} {\fixingtime_\bug^\developer \variable_{\epoch\developer\bug}}} , \label{eq:cost_function}
\end{align}
where $\fixingtime_\bug^\developer$ is the fixing cost associated with the experience of developer $\developer$ in addressing bug of type $\bug$. The postponement cost $\coefficient(\bug_\texttt{due}, \epoch)$ gives the incentive that a bug should be addressed as close as possible to its due date. We assume a linear cost for the postponement as follows:
\begin{align}
    \coefficient(\bug_\texttt{due}, \epoch) = \frac{\nepochs - \bug_\texttt{due}}{\nepochs}.
\end{align}

Using this linear function, we aim to have a postponement penalty if the due date is not reached. When we arrive at the due date, the cost gets 1, and after the due date has passed, it grows, imposing a higher cost for delaying overdue bugs. Attribute $\bug_\texttt{due}$ of the bugs is a function of $\epoch$, starting from the maximum assignment time, $\nduedate$, and reduced by one at each epoch if we decide to defer the bug. Moreover, $\nepochs$ denotes the total number of epochs, normalizing all the bug due dates. 


\subsubsection{Exogenous Information}
We have three types of exogenous information in our model. First, new bugs arrive randomly in the system following the underlying stochastic process. Their arrival times follow the exact distributions as the actual data. It might be different for the training and testing phases. We denote this exogenous information at epoch $\epoch$ as $\exoginfo^\texttt{bug}_\epoch$. It determines the number of newly reported bugs of each type to the system at the end of epoch $\epoch$. We do not have any presumption on the independence of bug arrival time in the system. Second, we may have a sudden change in the developers' schedule. That way, a developer may give short notice of being absent/present in the next epoch, which is expected in the real case scenarios, denoted by $\exoginfo^\texttt{sch-change}_\epoch$. That is, $\exoginfo^\texttt{sch-change}_\epoch$ realizes any unexpected change in the developers' schedule. Third, after assigning a bug to a developer, they may accept/reject fixing the bug based on their preference. The exploration parameter of the model, $\epsilon$, regulates the likelihood of order rejection in the system. 
We further discuss the impact of exploration versus exploitation in Section~\ref{sec:exploration}. 
By defining the exogenous information of declining to fix an assigned bug as $\exoginfo^\texttt{rejection}_\epoch$, the exogenous information arrival vector for epoch $\epoch \in \{1, \dots, \nepochs-1\}$ between decisions epochs $\epoch$ and $\epoch+1$ is denoted as $\exoginfo_\epoch = \big( \exoginfo^\texttt{bug}_\epoch, \exoginfo^\texttt{sch-change}_\epoch, \exoginfo^\texttt{rejection}_\epoch \big)$.

\subsubsection{Transition Function}
The transition function, shifting the system state forward in time, can be divided into two portions using post-decision states, as explained in Section~\ref{sec:ADP}. 
We define two transition functions as follows:
\begin{align}
    \StateSpace_{\epoch}^{\texttt{bug-post}}=\texttt{statepost}(\action_\epoch,\StateSpace_{\epoch}^{\texttt{bug}}) \nonumber\\
    \StateSpace_{\epoch}^{\texttt{dev-post}}=\texttt{statepost}(\action_\epoch,\StateSpace_{\epoch}^{\texttt{Dev}}) \label{eq:statepost}.
\end{align}

\begin{figure*}[!ht]
    \centering
    \includegraphics[width=.7\textwidth]{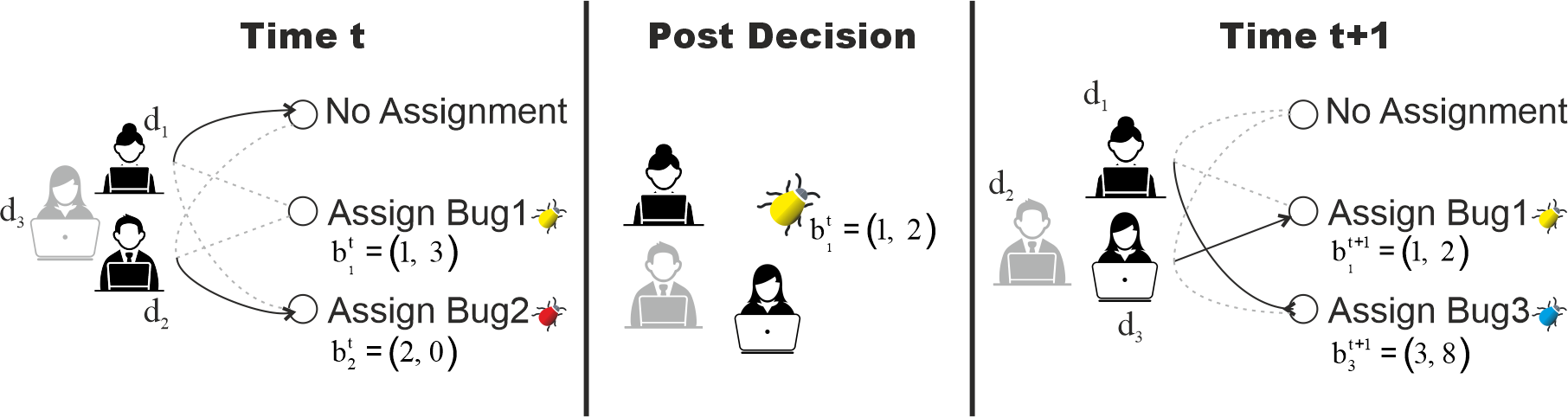}
    \caption{The transition of states in the bug assignment problem}
    \label{fig: motivating}
\end{figure*}

Given the current state of the bugs and developers, as well as the tuple of actions, we update the bug and developer states forward in time to their respective post-decision states. They describe the states of bugs and developers after they finish the defined actions but before exogenous information is introduced in the next time step. Following that, we define the second set of transition functions as
\begin{align}
    \StateSpace^\texttt{bug}_{\epoch+1}=\texttt{statenext}(\StateSpace^\texttt{bug-post}_\epoch,\exoginfo_\epoch) \nonumber\\
    \StateSpace^\texttt{Dev}_{\epoch+1}=\texttt{statenext}(\StateSpace^\texttt{dev-post}_\epoch,\exoginfo_\epoch) \label{eq:statenext}
\end{align}
that takes the bug and developer post-decision states, together with the exogenous information from the next time step, moving the system forward in time to the next set of states at epoch $\epoch+1$. Figure~\ref{fig: motivating} shows a typical transition of developers and bugs in the system. At epoch $\epoch$, two out of three developers are available. \texttt{ADPTriage} recommends assigning $\bug_2$ to $\developer_2$. Accordingly, in the post-decision state, the assigned developer $\developer_2$ becomes unavailable for the next $\fixingtime_\bug^\developer$ epochs, whereas the other developer remains available. The unassigned bug will move to the post-decision state while its $\bug_\texttt{due}$ attribute is reduced by 1. On the other hand, developer $\developer_3$ becomes available and is added back to the system. To move the post-decision state, we observe the exogenous variable. Based on the new information, bug $\bug_3$ is reported to the system, developer $\developer_2$ accepts to work on the assigned bug, and there is no last-minute change in the developers' schedules. Therefore, the system state at $\epoch+1$ includes two developers and two bugs, respectively.

Based on the aforementioned definition, we denote the number of bugs of type $\bug$ that are available after the assignment is made and the number of developers in the post-decision state as
\begin{align}
    \StateSpace_{\epoch(\bug_\LDA, \bug_\Due-1)}^{\texttt{bug-post}} & = \postponed_{\epoch(\bug_\LDA, \bug_\Due)} \label{eq:transition_bugs} \\
    \StateSpace_{\epoch(\developer_\Exp,\developer_\Sch=0)}^{\texttt{dev-post}} & = \slack_{\epoch(\developer_\Exp,\developer_\Sch=0)} + \StateSpace_{\epoch(\developer_\Exp,\developer_\Sch=1)}^\texttt{dev} \label{eq:transition_devs1}
\end{align}
where the number of bugs in the post-decision state depends only on the number of postponed bugs with their updated $\bug_\texttt{due}$ attribute. Moreover, the number of available developers with attribute $\developer$ in the post-decision state is equal to the number of unassigned developers of the same type and the ones whose $\developer_\texttt{sch}$ attribute was 1 and is going to be available in epoch $\epoch+1$ (i.e., their $\developer_\texttt{sch}$ will become 0).
We further make a transition from the post-decision state to the state at the next epoch, $\epoch+1$, which depends on the exogenous information. 

\subsubsection{ADP Formulation}\label{sec:adpformulation}
Even for medium-sized cases, the described dynamic problem has intractability in decision outcomes, states, and action spaces. This phenomenon is referred to as the ``three curses of dimensionality'', and the ADP framework is recommended as a solution technique for alleviating such complexity. We suggest a look-ahead strategy, \texttt{ADPTriage}, derived from value function approximation as a solution for bug triage. We leverage Linear Programming, Reinforcement Learning, and Natural Language Processing to construct our \texttt{ADPTriage}. 

\paragraph{Value Function Approximation} We employ a linear approximation for the post-decision state value function as follows:
\begin{align}
   \bar{\valuefunction}_\epoch^{\texttt{Post}} (\StateSpace_\epoch^{\texttt{bug-post}},\StateSpace_\epoch^{\texttt{dev-post}}) = & \Big(\sum_{\bug \in \BugSet} \bar{\mathit{v}}_{\epoch\bug}^\texttt{Post}\StateSpace_{\epoch\bug}^{\texttt{bug-post}} \nonumber + \\
   & \sum_{\developer \in \DeveloperSet} \bar{\mathit{v}}_{\epoch\developer}^\texttt{Post}\StateSpace_{\epoch\developer}^\texttt{dev-post}\label{eq:val_approx} \Big). 
\end{align}

That is, we decompose the joint value function $\bar{\valuefunction}_\epoch^{\texttt{Post}} (\StateSpace_\epoch^{\texttt{bug-post}},\StateSpace_\epoch^{\texttt{dev-post}})$ into the value functions of individual bug and developer types since our system correlates rewards with available bugs' and developers' attributes. We sum over all postponed bugs in the first portion of Equation~\eqref{eq:val_approx} with their expected future value estimate of $\bar{\mathit{v}}_{\epoch\bug}^\texttt{Post}$. In the second part of Equation~\eqref{eq:val_approx}, we take the sum of all developers with attribute $\developer$ that are available at epoch $\epoch$ and multiply it with its future value estimate using the corresponding developer's value function approximations. Equation~\eqref{eq:val_approx} approximates the corresponding value function of our comprehensive system state of bugs and developers at time $\epoch$.

\paragraph{ADP Policy}
In our bug triage problem, the objective is to minimize the expected cost over the project horizon. The cost is associated with the suitability of the developers and is estimated by their bug fixing times. Therefore, we solve the following linear programming model to find the decisions at epoch $\epoch$:
\begin{equation}
    \begin{aligned}
        \valuefunction_\epoch(\state_\epoch) = & \min_{\action_\epoch \in \ActionSet_\epoch (\state_\epoch)}\  {\big(\text{Cost}(\state_\epoch, \action_\epoch) + \bar{\valuefunction}_\epoch^{\texttt{Post}} (\StateSpace_\epoch^{\texttt{bug-post}},\StateSpace_\epoch^{\texttt{dev-post}}) \big)} \\
        \text{s.t.}\ \ & \eqref{eq:assigned_devs}, \eqref{eq:assigned_bugs}\\
        & \variable_{\epoch\developer\bug}, \postponed_{\epoch\bug}, \slack_{\epoch\developer} \in \mathbb{R}. \label{eq:ADP_IP}
    \end{aligned}
\end{equation}

We employ the forward-pass ADP algorithm as shown in Algorithm~\ref{alg:ADPTriage}.

\begin{algorithm}[!ht]
    \textbf{Step 0:} Initialize post-decision values $(\bar{\mathit{v}}_{\epoch\bug}^{\texttt{Post},0}, \bar{\mathit{v}}_{\epoch\developer}^{\texttt{Post},0})$, state $\state_1^0$, and step size $\stepsize^0$\\
	\BlankLine
	\textbf{Step 1:} \For{$\iteration \in \{1,\hdots,\niteration\}$}{
	    \textbf{Step 2:} Choose a sample path $\big(\hat{\exoginfo}^\iteration_\epoch\big)_{\epoch \in \EpochSet}$ for exogenous information
	    
    	\textbf{Step 3:} \For{$\epoch \in \{1,\hdots,\nepochs\}$}{
            \textbf{Step 3.a:} Solve the ADP equation in Equation~\eqref{eq:ADP_IP}, find the optimal decisions $\action^\iteration_\epoch$.
            
            \textbf{Step 3.b:} Compute the marginal values of Constraints \eqref{eq:assigned_devs} and \eqref{eq:assigned_bugs} to estimate $\bar{\mathit{v}}_{\epoch\developer}^{\texttt{Post},\iteration}$ and $\bar{\mathit{v}}_{\epoch\bug}^{\texttt{Post},\iteration}$, respectively.
            
            \textbf{Step 3.c:} Update post-decision value function approximation of the previous epoch, $\bar{\mathit{v}}_{\epoch-1,\developer}^{\texttt{Post},\iteration-1}$ and $\bar{\mathit{v}}_{\epoch-1,\bug}^{\texttt{Post},\iteration-1}$ using Equation~\eqref{eq:alpha update}.
            
            \textbf{Step 3.d:} Move to the new state $\state_{\epoch+1}^\iteration$ after taking action $\action_\epoch^\iteration$ and observing the exogenous information $\exoginfo_\epoch^\iteration$ via Equations \eqref{eq:statepost} and \eqref{eq:statenext}.
        }
	}
	\textbf{Step 4:} Return the final approximation values of post-decision states, $\bar{\mathit{v}}^{\texttt{Post},\niteration} =  (\bar{\mathit{v}}_{\epoch\bug}^{\texttt{Post},\niteration}, \bar{\mathit{v}}_{\epoch\developer}^{\texttt{Post},\niteration})$, for the future online decision making
	\caption{Forward-pass \texttt{ADPTriage}}
	\label{alg:ADPTriage}
\end{algorithm}

We initialize the post-decision values by the postponement penalty (e.g., the maximum fixing time) and start the policy learning iterations. In each iteration, we first choose a sample path of length $\nepochs$ (i.e., project horizon) for exogenous information (i.e., developers' schedules/availabilities and bug arrivals). Then, we follow the random path to find the optimal decisions by solving Equation~\eqref{eq:ADP_IP}. After finding the optimal decisions, we estimate the future values of post-decision states. Two marginal values require to be computed: (1) the cost of assigning a bug to a developer or using their experience for more suitable bugs in upcoming epochs, and (2) the marginal values of the cost of postponing a bug or assigning now. These two values can be obtained through computing the duals of the Constraints \eqref{eq:assigned_devs} and \eqref{eq:assigned_bugs}. 

After approximating the post-decision state values, we update the post-decision value of the previous epoch for all elements in $\state_\epoch$ using

\begin{equation}
    \begin{aligned}
        \bar{\mathit{v}}_{\epoch-1\developer}^{\texttt{Post},\iteration} = & (1-\stepsize^\iteration)\bar{\mathit{v}}_{\epoch-1\developer}^{\texttt{Post},\iteration-1} + \stepsize^\iteration \bar{\mathit{v}}_{\epoch-1\developer}^{\texttt{Post},\iteration} \\
        \bar{\mathit{v}}_{\epoch-1\bug}^{\texttt{Post},\iteration} = & (1-\stepsize^\iteration) \bar{\mathit{v}}_{\epoch-1\bug}^{\texttt{Post},\iteration-1} + \stepsize^\iteration \bar{\mathit{v}}_{\epoch-1\bug}^{\texttt{Post},\iteration}
        \label{eq:alpha update}
    \end{aligned}
\end{equation}
where $\stepsize^\iteration$ is the desired step size at iteration $\iteration$ (e.g., 0.5 for the constant case).

We continue to the next epoch after random information realization $\exoginfo_\epoch^\iteration$. By repeating the process for $\niteration$ iterations, the model returns the final post-decision value estimates, $\bar{\mathit{v}}^{\texttt{Post},\niteration}$, which can be employed for the online bug triage assignment. For the online bug assignment, we fix those obtained $\bar{\mathit{v}}^{\texttt{Post},\niteration}$ estimates, and use those to generate a policy for any given state of the system. 

\subsubsection{Enhancements}
Updating the post-decision state values is crucial in finding the optimal solution. In the basic ADP, we use a constant value (e.g., 0.5) for the step size $\stepsize$. However, in the first iterations, our knowledge about the future values of the current state is inadequate. As we explore and revisit states, we better understand the long-run costs of a decision made. Therefore, we may adjust updating the step size according to the number of iterations, the number of times we visited a state or the difference between the current state value and its previous one. Here, we explore two enhancements for updating step size compared to a basic constant step size as follows. 

\paragraph{Harmonic step size update} The harmonic step size at iteration $\iteration$ is defined as
\begin{align}
    \stepsize^\iteration = \max \Big\{ \frac{\lambdahram}{\lambdahram + \iteration - 1}, \stepsize^0 \Big\}
\end{align}
where we set $\lambdahram$ to 25, and $\stepsize^0$ to 0.05, similar to \citet{mes2017}. It is an arithmetically declining sequence with the limit point of $\stepsize^0$. Larger values for $\lambdahram$ slow down the decline and enhance the learning process.

\paragraph{BAKF step size update} The bias-adjusted Kalman filter (BAKF), which is bounded by $\nicefrac{1}{\iteration}$ is an efficient way of learning and converging the value function. \citet{powell2007approximate} shows that it leads to a faster convergence compared to other step size updates. The BAKF step size balances observation variation and transient bias. As a result, it uses smaller step sizes for value function observations with a large variation and low bias, whereas it suggests larger step sizes for value function observations with a small variation and high bias. 

\begin{algorithm}[!ht]
    \textbf{Step 0:} Initialize $\bar{\BAKFestimate}^0$, $\stepsize^0$, $\BAKFappbias^0$, $\BAKFerrorstep^0$, $\bar{\BAKFerrorstep}$ and $\BAKFsqrpred^0$.
    
	\BlankLine
	\textbf{Step 1:} Obtain the new observation $\bar{\mathit{v}}^{\texttt{Post},\iteration}$. 
	
	\textbf{Step 2:} Update the following parameters accordingly:
	\begin{align*}
	    \BAKFerrorstep^\iteration & = \frac{\BAKFerrorstep^{\iteration-1}}{1+\BAKFerrorstep^{\iteration-1}-\bar{\BAKFerrorstep}} \\
        \bar{\BAKFappbias}^\iteration & = (1-\BAKFerrorstep^{\iteration})\bar{\BAKFappbias}^{\iteration-1} + \BAKFerrorstep^{\iteration} \big( \bar{\mathit{v}}^{\texttt{Post},\iteration} - \bar{\mathit{v}}^{\texttt{Post},\iteration-1} \big)\\
        \bar{\BAKFsqrpred}^\iteration & = (1-\BAKFerrorstep^{\iteration})\bar{\BAKFsqrpred}^{\iteration-1} + \BAKFerrorstep^{\iteration} \big( \bar{\mathit{v}}^{\texttt{Post},\iteration} - \bar{\mathit{v}}^{\texttt{Post},\iteration-1} \big)^2 \\
        (\bar{\BAKFnoise}^\iteration)^2 &= \frac{\bar{\BAKFsqrpred}^\iteration - (\bar{\BAKFappbias}^\iteration)^2}{1+\bar{\BAKFvariance}^{\iteration-1}}
	\end{align*}
	
	\textbf{Step 3:} Evaluate the step sizes for the current iteration (if $\iteration>1$).
	\begin{align*}
	    \stepsize^\iteration & = 1- \frac{(\bar{\BAKFnoise}^\iteration)^2}{\bar{\BAKFsqrpred}^\iteration} 
	\end{align*}	

	\textbf{Step 3:} Update the coefficient for the variance of the smoothed estimate.
	\begin{align*}
	    \bar{\BAKFvariance}^{\iteration} & = 
	    \begin{cases}
	        (\stepsize^\iteration)^2 & \text{if}~n = 1\\
	        (1-\stepsize^\iteration)^2\bar{\BAKFvariance}^{\iteration-1} + (\stepsize^\iteration)^2 & \text{if}~n > 1
	    \end{cases}
	\end{align*}	

	\textbf{Step 4:} Smooth the value function estimate using the obtained step size $\stepsize^n$ and Equations~\eqref{eq:alpha update}.
	\caption{BAKF - Optimal Step size Algorithm}
	\label{alg:BAKF}
\end{algorithm}

Algorithm~\ref{alg:BAKF} shows the step size estimation via BAKF algorithm. In Step 0, we initialize the parameters of the models as follows: $\bar{\BAKFestimate}^0=0$, $\stepsize^0=1$, $\BAKFappbias^0=0$, $\BAKFerrorstep^0=0.01$, $\bar{\BAKFerrorstep}=0.2$ and $\BAKFsqrpred^0=0$.

\subsection{Myopic Policy}
As the baseline, we design an optimization-based myopic policy for the bug triage problem based on intuition regarding bug assignment. This benchmark does not take into account the future arrival of bugs or the potential schedule change of developers. The term ``myopic'' means that it ignores the cost of decisions made farther down the stream. As a result, it is comparable to the ADP method but without the learning component. We set the coefficient of postponement, $\coefficient(\bug_\texttt{due}, \epoch)$, in Equation~\eqref{eq:cost_function} to a large number (e.g., the Big-M) to force the myopic approach to exploit the full capacity of the developers at each epoch without considering the possible merits of postponement. Hence, the myopic approach is converted into a knapsack problem in which we aim to reduce the fixing time while imposing constraints on developers' schedules and capacities. This approach is first introduced by~\citet{kashiwa2020} as the release-aware bug triage method (RABT). It is shown that their approach outperforms other baselines, e.g., content-based recommendation and cost-aware bug triage~\citep{kashiwa2020,jahanshahi2022sdabt}. As a result, we select their IP solution as the myopic baseline to be compared with our MDP solution. 

\subsection{Design of the Experiment}
We implement our approaches by following the three steps listed below:
\begin{itemize}
    \item[(1)] We leverage the Wayback Machine, introduced by~\citet{jahanshahi2020Wayback}, for the training period. Using the tool, we extract bugs` LDA categories and assignment times. We need the information to understand the distribution of assignment deadlines and also cluster bugs to certain categories according to their textual information.
    
    \item[(2)] We run the Wayback Machine once more to extract the distribution of bug arrival times per bug category and developers' availability. The expertise of each developer in fixing different bug categories is also determined. Accordingly, we obtain a large picture of the environment, which is essential to formulate an MDP problem.
    
    \item[(3)] With bugs' and developers' information at hand, we run the \texttt{ADPTriage} according to Algorithm~\ref{alg:ADPTriage}. Similar to the previous two steps, it is implemented only for the training period. The output of this step is an optimal ADP policy that can be utilized during the testing phase.
\end{itemize}

After finding the optimal policy, we employ this policy for the testing phase and report the model performance based on the statistics collected during the testing phase. We note that the LDA categories of the bugs in the testing phase are estimated based on the model we obtained during the training phase. We do not have any retraining afterwards.

\section{Results} \label{sec:results3}

We conduct a detailed numerical study to compare \texttt{ADPTriage} and its enhancements against the myopic approach.
We primarily focus on exploring whether \texttt{ADPTriage} postpones a bug to find a suitable developer in future timestamps.
We utilize Gurobi 9.5 to solve the linear programming models and implement all of the algorithms in Python.

\subsection{Performance Analysis}
\begin{figure*}[!ht]
  \begin{center}
        \subfloat[{\scshape{EclipseJDT}} \label{fig:fixing_time_EclipseJDT}]
        {\includegraphics[width=0.33\textwidth]{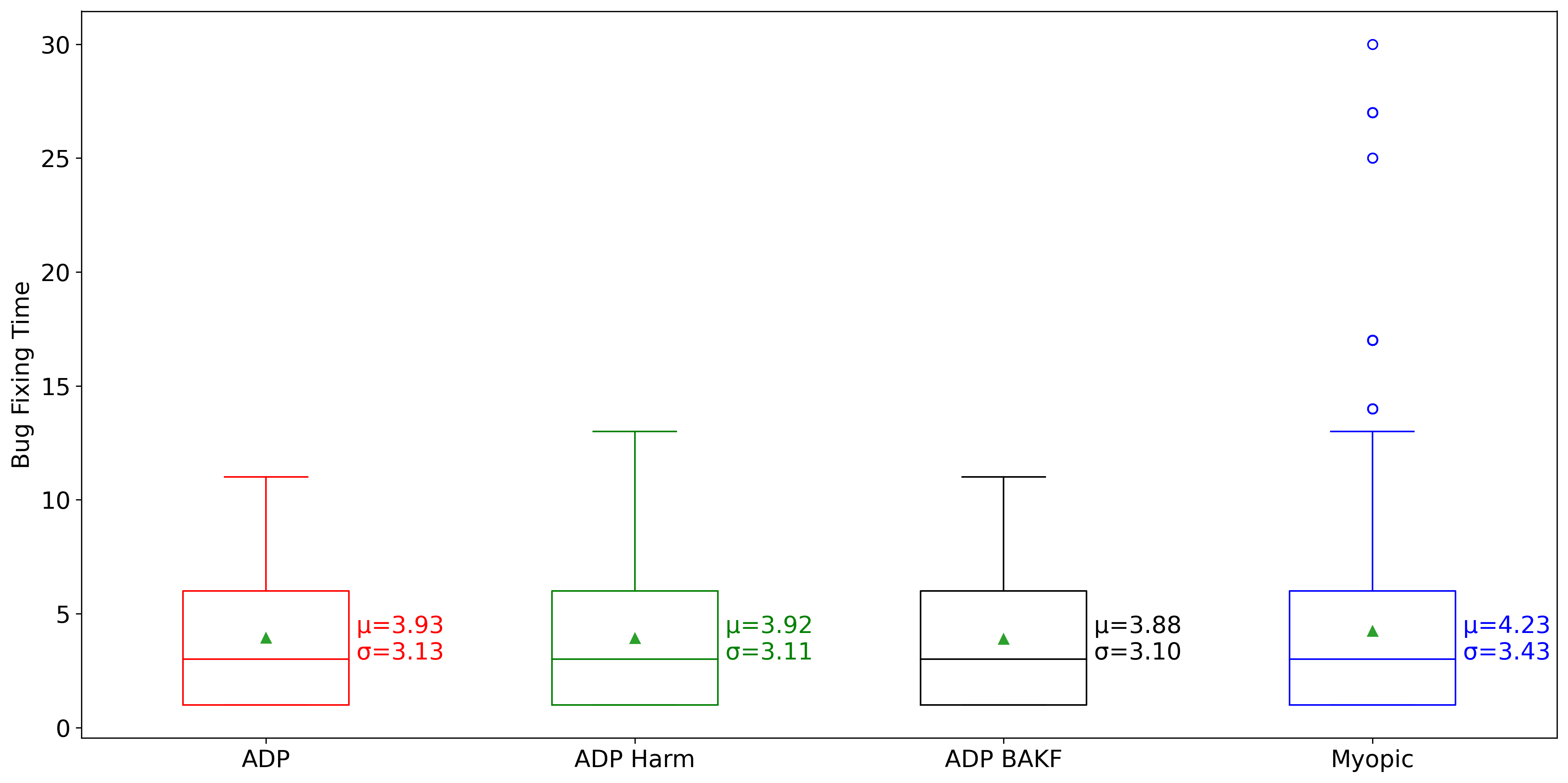}}~
        \subfloat[{\scshape{GCC}} \label{fig:fixing_time_GCC}]
        {\includegraphics[width=0.33\textwidth]{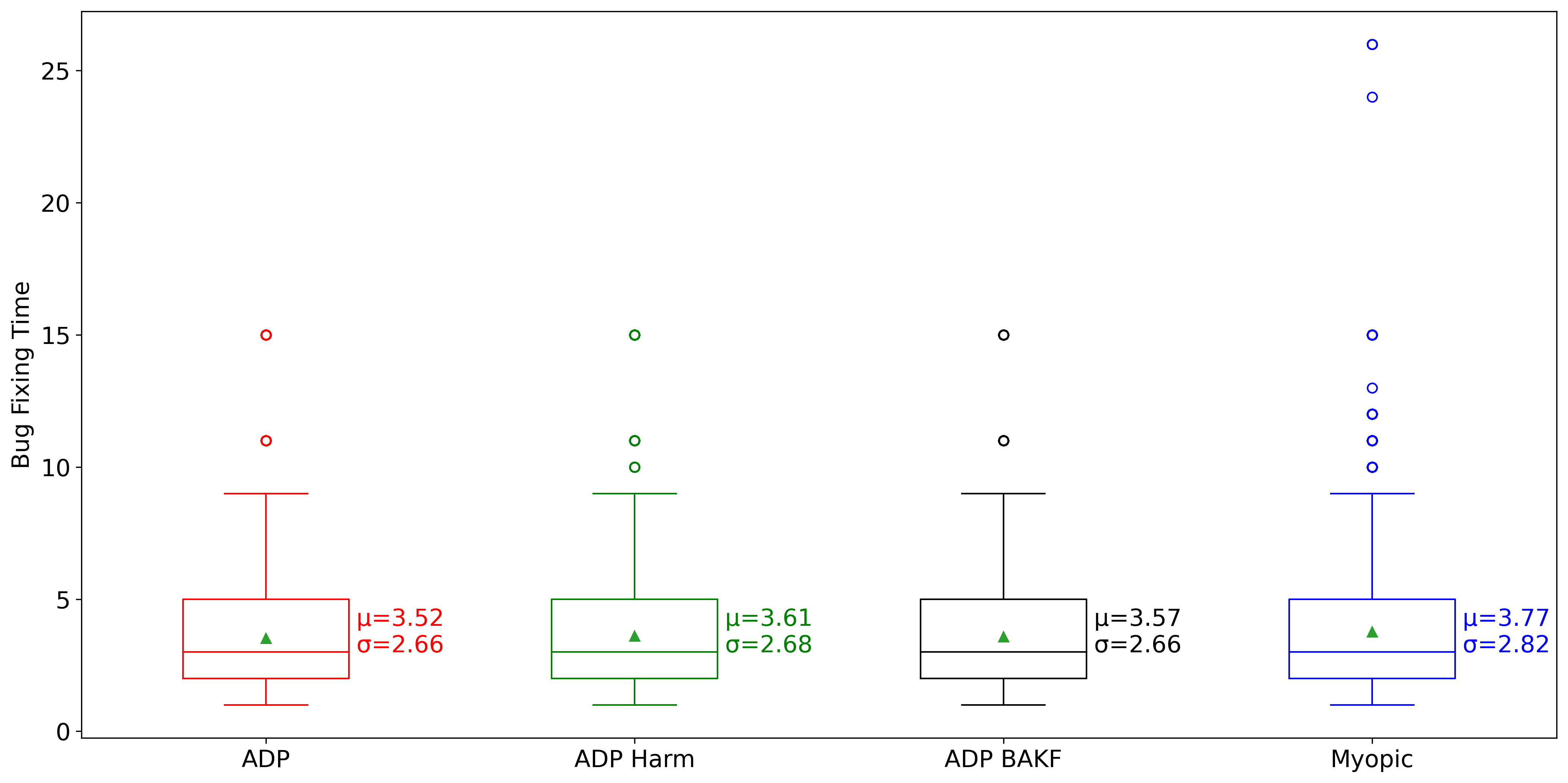}}
        \subfloat[{\scshape{Mozilla}} \label{fig:fixing_time_Mozilla}]
        {\includegraphics[width=0.33\textwidth]{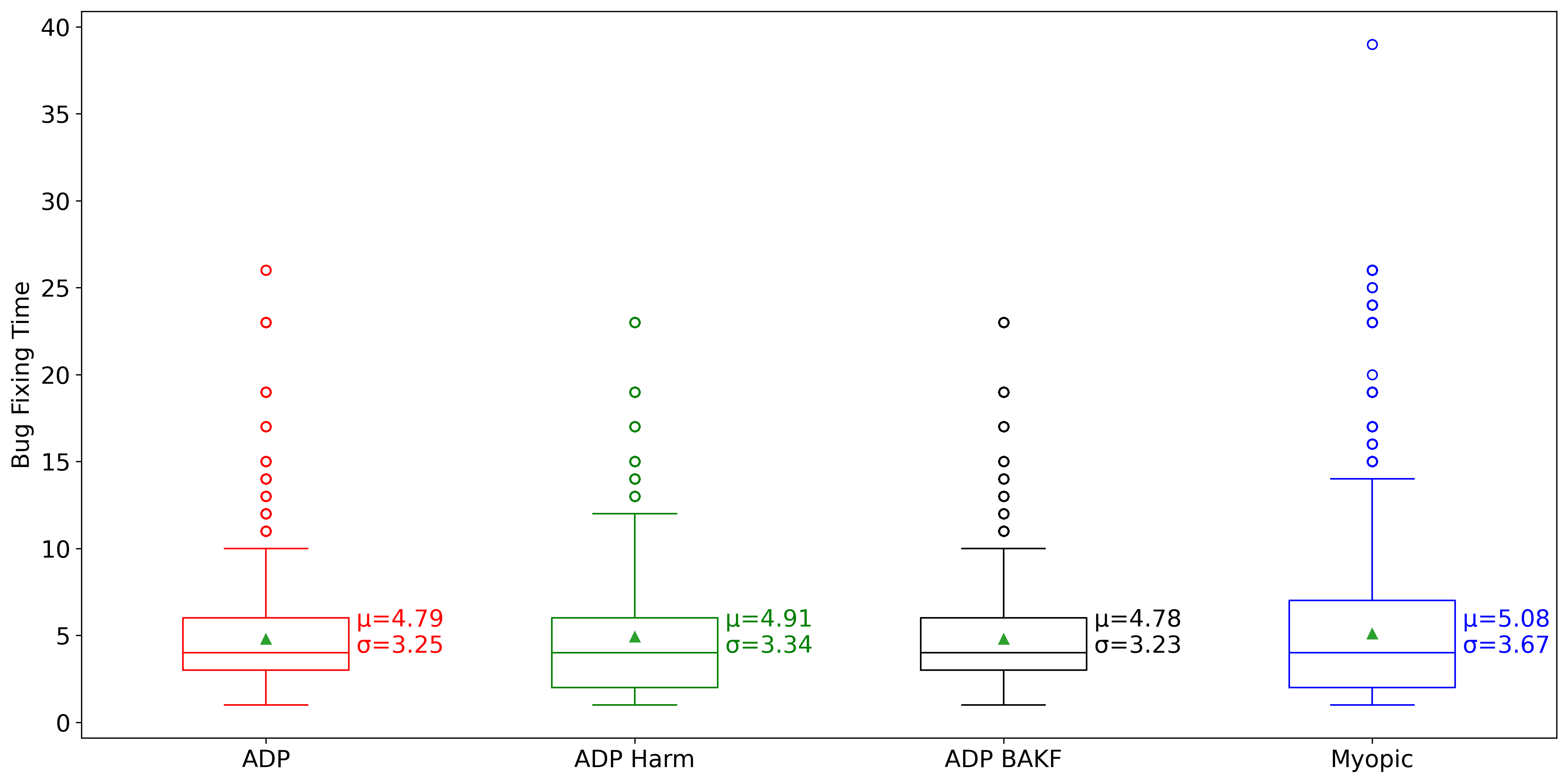}}
        \caption{Fixing time of the bugs during the testing phases}
        \label{fig:fixing_times}
  \end{center}
\end{figure*}

By defining the expected time to assign a bug, we aim to find a policy that optimizes the assignment time such that the most suitable developer (i.e., the developer with the shortest fixing time) is assigned to a bug. Therefore, by using the \textit{bug fixing time} as a proxy for the suitability of a developer, the optimal policy requires shortening the fixing time through proper assignment timing. Figure~\ref{fig:fixing_times} shows the boxplot of the bug fixing time distribution for all algorithms. We observe an almost 10\% reduction in the fixing times of the bugs by assigning the developers via \texttt{ADPTriage} with BAKF update compared to the myopic one. We statistically test this observation to see whether the difference is significant~\citep{Garcia2010}. Friedman Aligned Ranks as a non-parametric test is selected to compare the algorithms. Given the same set of fixed bugs, the $p$-value of the test for the bug fixing time is equal to $1.05e-9$, $8.85e-23$, and $3.08e-17$ for {\scshape{EclipseJDT}}, {\scshape{GCC}}, and {\scshape{Mozilla}}, respectively. Hence, we conclude that differences in algorithms' fixing times are statistically significant (for $\alpha = 0.05$).

A post hoc analysis is employed to investigate the source of the discrepancy. The post hoc approach evaluates the pairwise differences between all algorithms in terms of the average ranking of absolute difference. It enables us to compare two models side by side~\citep{demvsar2006}. Table~\ref{tab:comp_algorithms3} shows the $p$-values of pairwise comparisons using the Nemenyi post hoc test after finding the significance in the Friedman test. As the $p$-values of the myopic approach versus either of the ADP versions are much smaller than $\alpha = 0.05$ for all projects, we conclude that the fixing time (i.e., the suitability of the developers) for the ADP algorithms is significantly shorter than that of the myopic approach.

\begin{table*}[!ht]
\centering
\caption{Comparison of different algorithms using Nemenyi post hoc test ($p$-values are presented in the table.).\label{tab:comp_algorithms3}}%
\resizebox{\textwidth}{!}{
    \begin{tabular}{l|llll|llll|llll}
        \toprule
     & \multicolumn{4}{c|}{\textbf{{\scshape{EclipseJDT}}}} & \multicolumn{4}{c|}{\textbf{{\scshape{GCC}}}} & \multicolumn{4}{c}{\textbf{{\scshape{Mozilla}}}} \\
        & \textbf{Myopic} & \textbf{ADP} & \textbf{ADP Harm} & \textbf{ADP BAKF} & \textbf{Myopic} & \textbf{ADP} & \textbf{ADP Harm} & \textbf{ADP BAKF} & \textbf{Myopic} & \textbf{ADP} & \textbf{ADP Harm} & \textbf{ADP BAKF} \\
          \midrule
        \textbf{Myopic}   & 1.000 & 0.001***\textsuperscript{a} & 0.003** & 0.001*** & 1.000 & 0.001*** & 0.001*** & 0.001*** & 1.000 & 0.001*** & 0.001*** & 0.001*** \\
        \textbf{ADP}      &  & 1.000 & 0.900 & 0.900 &  & 1.000 & 0.072\textsuperscript{.} & 0.275 & & 1.000 & 0.004* & 0.900 \\
        \textbf{ADP Harm} &  &       & 1.000 & 0.806 &  &  & 1.000 & 0.900 &  &  & 1.000 & 0.005** \\
        \textbf{ADP BAKF} &  &       &       & 1.000 &  &  &  & 1.000 &  &  &  & 1.000  \\
        \bottomrule
        \multicolumn{13}{l}{\begin{tabular}[c]{@{}l@{}} {\footnotesize \textsuperscript{a} Significance codes: *** $p < 0.001$ ** $p < 0.01$ * $p < 0.05$ \textsuperscript{.} $p < 0.1$} \end{tabular}} \\
    \end{tabular}
}
\end{table*}

\begin{figure*}[!ht]
  \begin{center}
        \subfloat[{\scshape{EclipseJDT}} \label{fig:due_date_EclipseJDT}]
        {\includegraphics[width=0.33\textwidth]{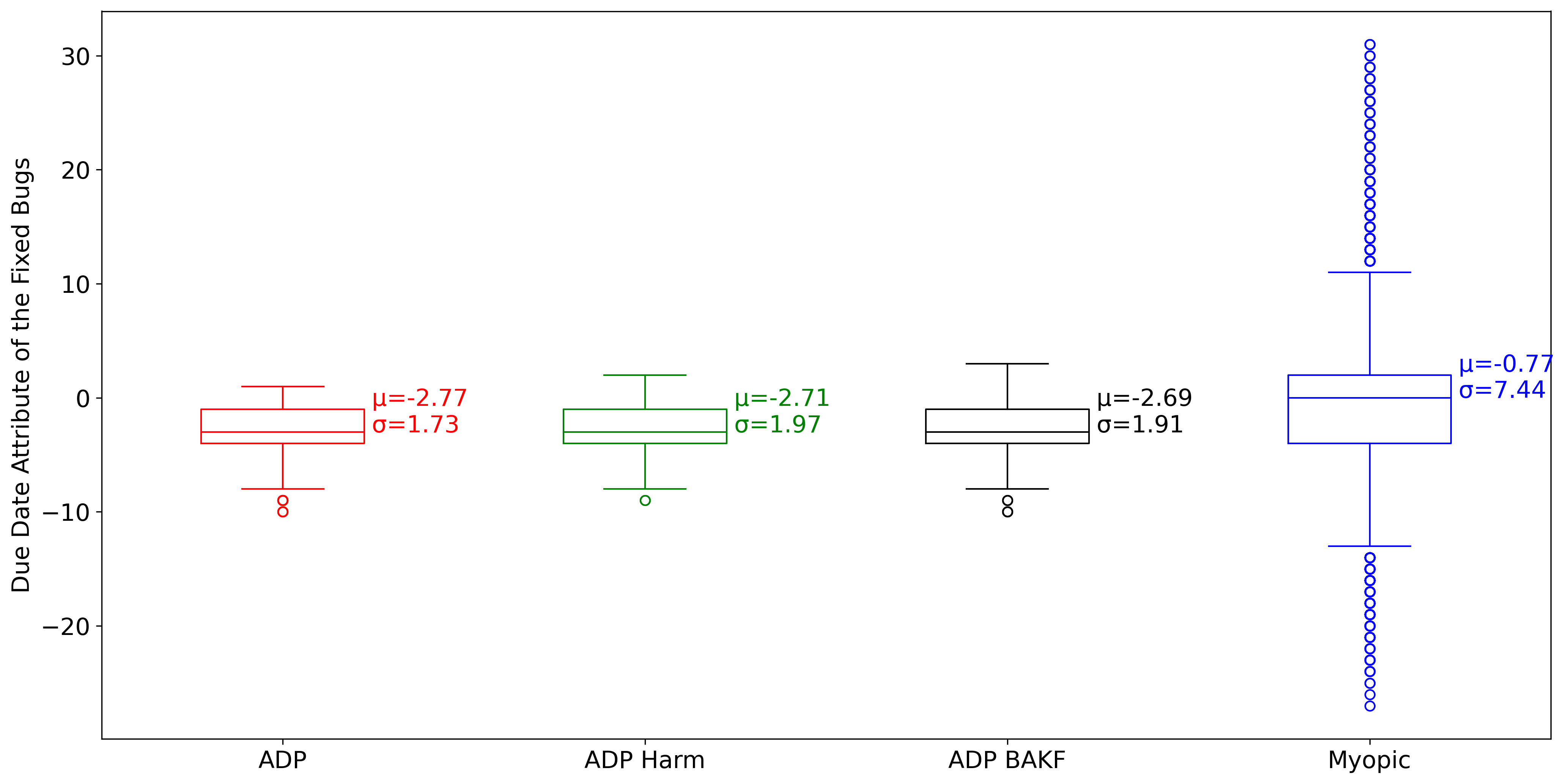}}~
        \subfloat[{\scshape{GCC}} \label{fig:due_date_GCC}]
        {\includegraphics[width=0.33\textwidth]{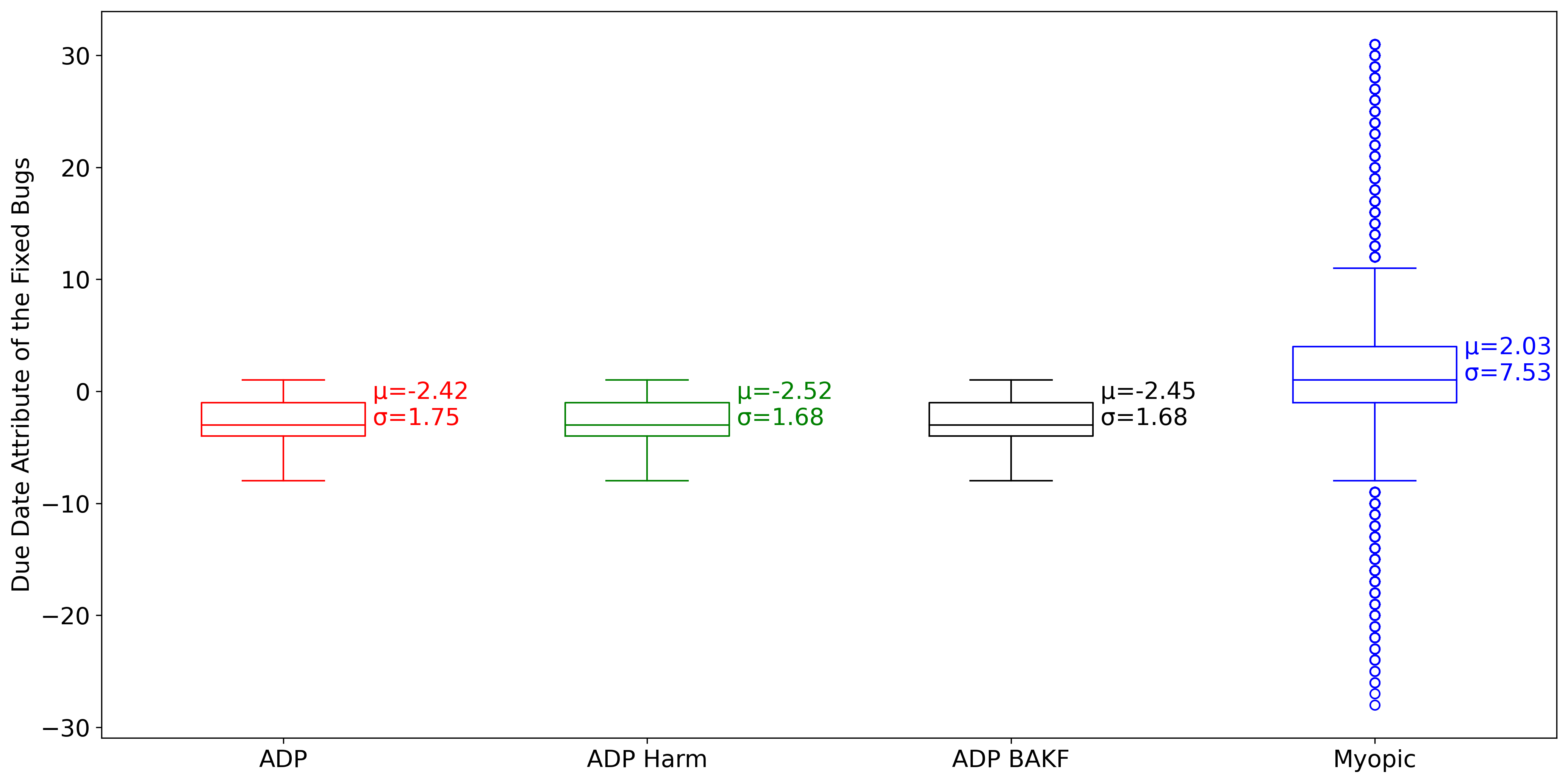}}
        \subfloat[{\scshape{Mozilla}} \label{fig:due_date_Mozilla}]
        {\includegraphics[width=0.33\textwidth]{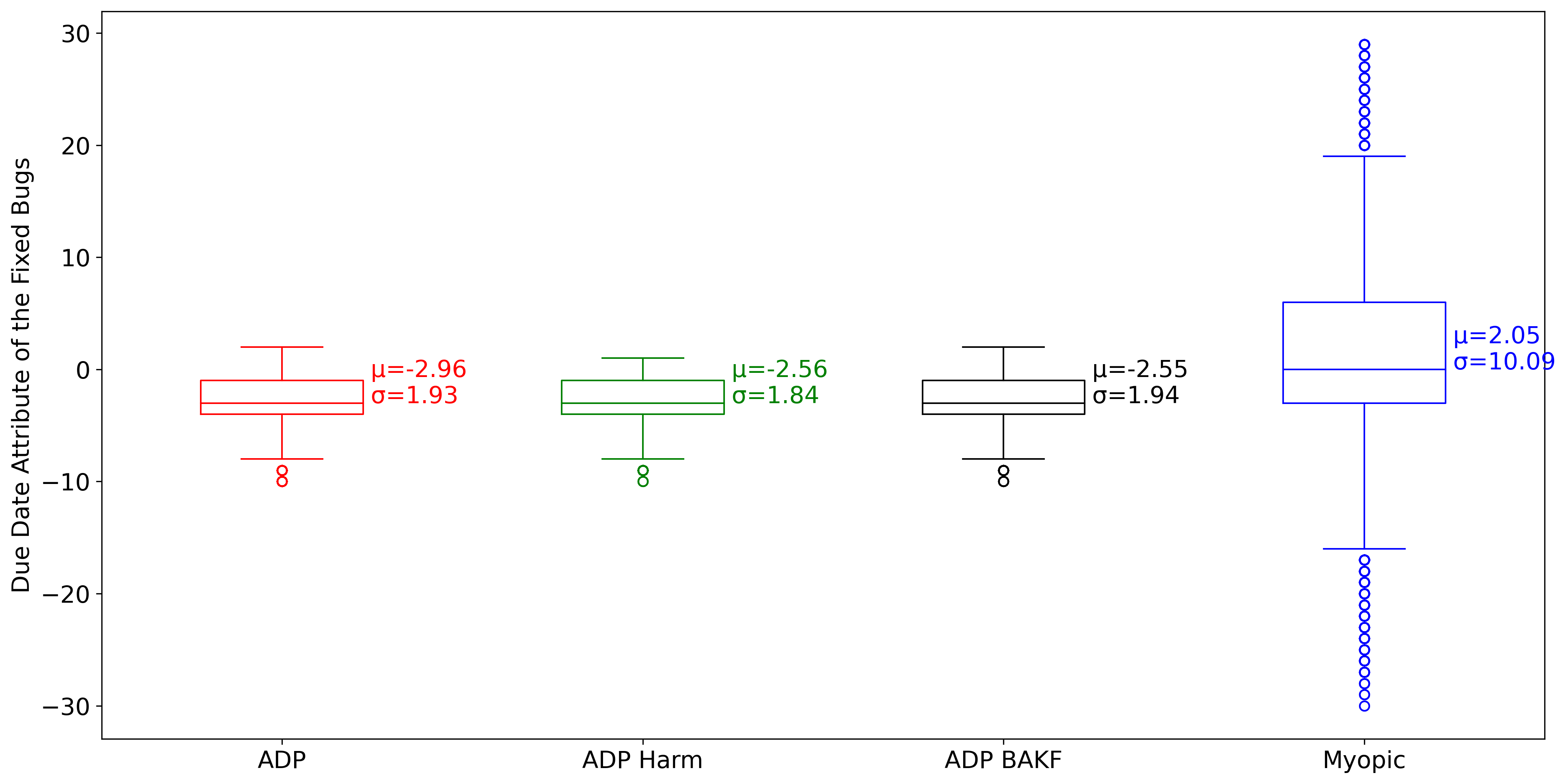}}
        \caption{The distribution of due dates of the bugs}
        \label{fig:due_date}      
  \end{center}
\end{figure*}

The intuitive interpretation for the significantly shorter bug fixing times of the ADP algorithms is that the ADP-based policy postpones bugs to find a better developer. Now, the important question is how much does the model sacrifice the timely fixing to achieve a better triage? Figure~\ref{fig:due_date} shows the distribution of the due date attribute of the bugs ($\bug_\Due$) in the assignment time. It indicates the number of days we had until on-time assignment. In other words, negative values correspond to late assignments, and positive values relate to the early ones. On average, \texttt{ADPTriage} approaches defer fixing the bugs by less than 2 days for {\scshape{EclipseJDT}} and almost 4 days for {\scshape{GCC}} and {\scshape{Mozilla}}  to accomplish its objective of assigning them to a better developer. ADP takes advantage of the presence of a more suitable developer in upcoming epochs to postpone bugs, matching them with the most cost-effective developer. Nonetheless, the average extra postponement seems small compared to the project horizon and deadline distribution. To summarize, ADP can achieve considerable cost reductions compared to the myopic approach for about two to four more periods of deferral. This insight is crucial for the triagers, who may be interested in the magnitude of ADP's postponements. We also observe a much greater variance in due dates of assigned bugs of the myopic algorithm than that of \texttt{ADPTriage}. Specifically, the BAKF version of ADP tends to triage bugs as close as possible to their due dates (i.e., close to 0). It indicates that the obtained policy of \texttt{ADPTriage} is more reliable in terms of punctual addressing of the bugs. 

We define the most suitable person to address a bug as the one with the shortest fixing time, i.e., expert enough to handle it as fast as possible. Hence, we determine top-$k$ developers. It indicates whether the proposed developer by a method is among top-$k$ developers in terms of the fastest fixing time. Table~\ref{tab: accuracy} shows the models' accuracy in terms of assigning the bugs to the appropriate developers. ADP algorithms demonstrate enhancement in accuracy even though they tend to postpone the bugs. It indicates that these postponements are made to find better developers rather than arbitrary deferrals. The improvement here may seem small; however, we note that according to the developers' schedules and unpredictable availabilities, it proves the proper learning process of the ADP.

\begin{table*}[!ht]
\centering
\caption{The accuracy of the models in assigning the bugs to the most suitable developers.\label{tab: accuracy}}%
\resizebox{.7\textwidth}{!}{
    \begin{tabular}{lrrr|rrr|rrr}
    \toprule
     & \multicolumn{3}{c|}{\textbf{{\scshape{EclipseJDT}}}} & \multicolumn{3}{c|}{\textbf{{\scshape{GCC}}}} & \multicolumn{3}{c}{\textbf{{\scshape{Mozilla}}}} \\
     & \textbf{top-1} & \textbf{top-3} & \textbf{top-5} & \textbf{top-1} & \textbf{top-3} & \textbf{top-5} & \textbf{top-1} & \textbf{top-3} & \textbf{top-5}\\
     \midrule
    \textbf{Myopic} & 27.9 & 62.1 & 79.2 & 24.4 & 49.1 & 56.9 & 5.6 & 16.7 & 21.2 \\
    \textbf{ADP} & 28.2 & 66.7 & 82.5 & \textbf{26.1} & \textbf{51.9} & \textbf{59.8} & \textbf{6.8} & \textbf{21.1} & 24.5 \\
    \textbf{ADP Harm} & 28.6 & 66.7 & \textbf{82.7} & 25.6 & 49.7 & 58.0 & 6.0 & 19.7 & \textbf{24.9} \\
    \textbf{ADP BAKF} & \textbf{29.0} & \textbf{66.9} & 82.4 & 25.5 & 50.5 & 58.5 & 6.2 & 19.7 & 24.2 \\
    \bottomrule
    \end{tabular}
}
\end{table*}

\subsection{Convergence}~\label{sec: convergence}
\begin{figure*}[!ht]
  \begin{center}
        \subfloat[Realized rewards - {\scshape{EclipseJDT}} \label{fig:discounted_reward}]
        {\includegraphics[width=0.5\textwidth]{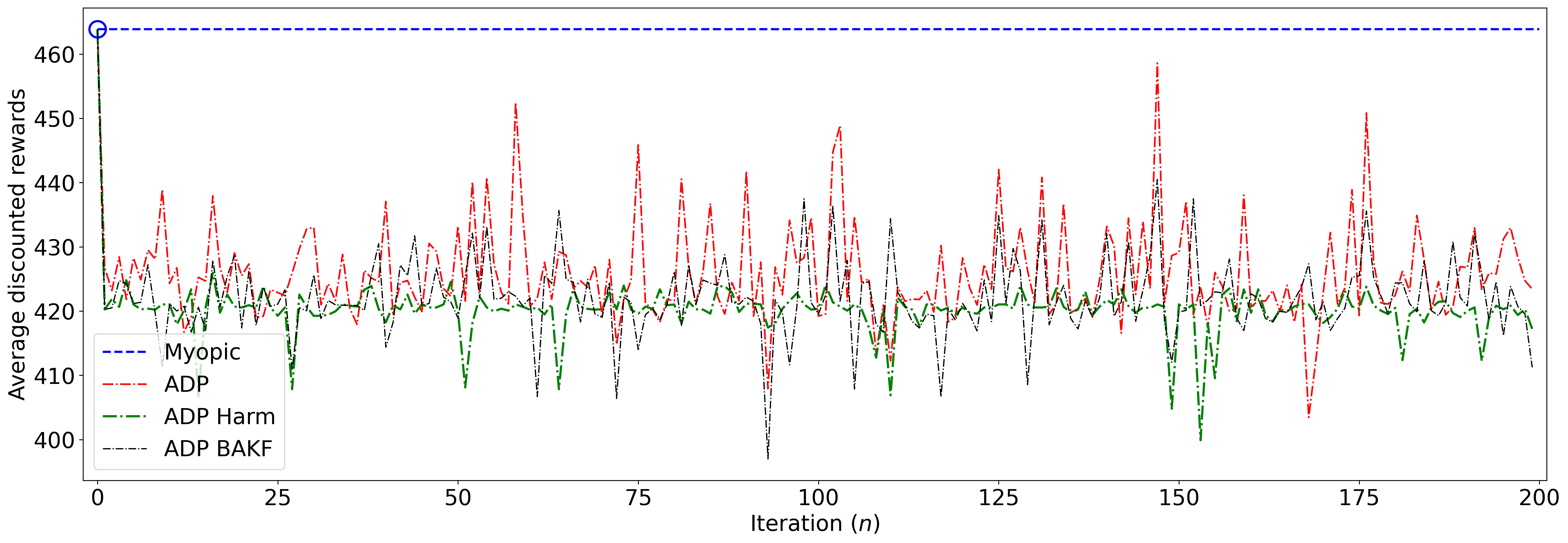}}~
        \subfloat[$\bar{\valuefunction}_0(\state_0)$ Estimate for bug type 1 and due date of -1  - {\scshape{EclipseJDT}}\label{fig:estimated_value_post}]
        {\includegraphics[width=0.5\textwidth]{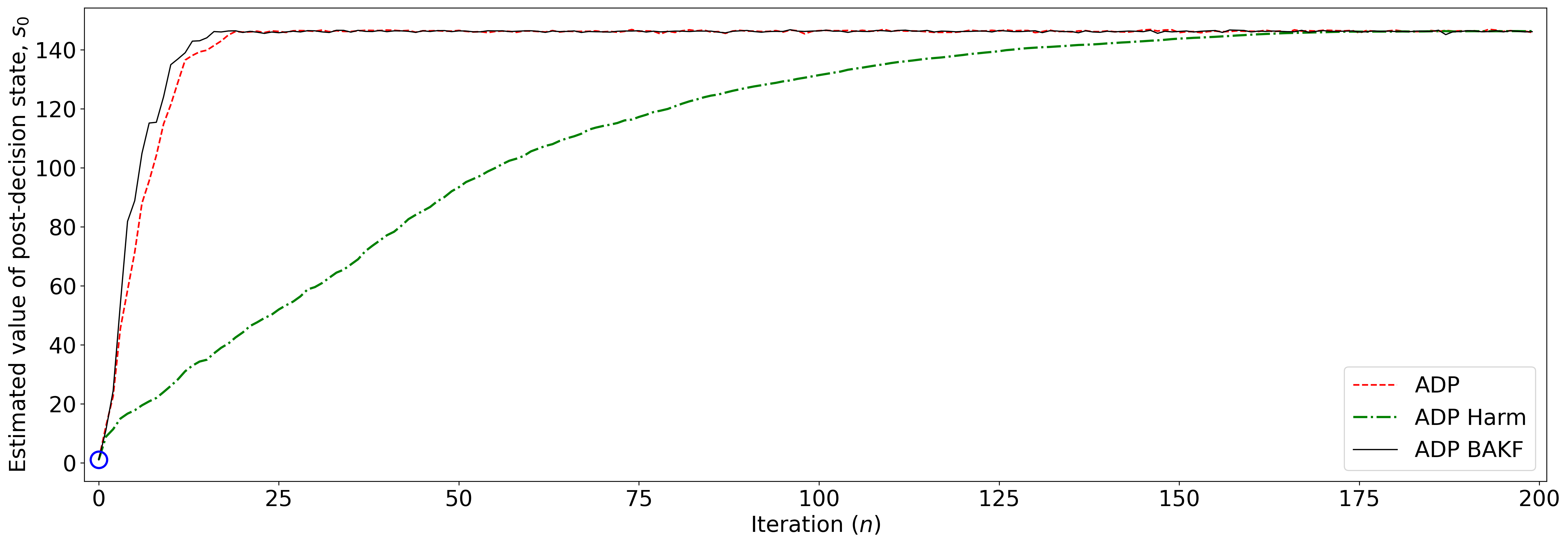}}\\
        \subfloat[Realized rewards - {\scshape{GCC}} \label{fig:discounted_reward_GCC}]
        {\includegraphics[width=0.5\textwidth]{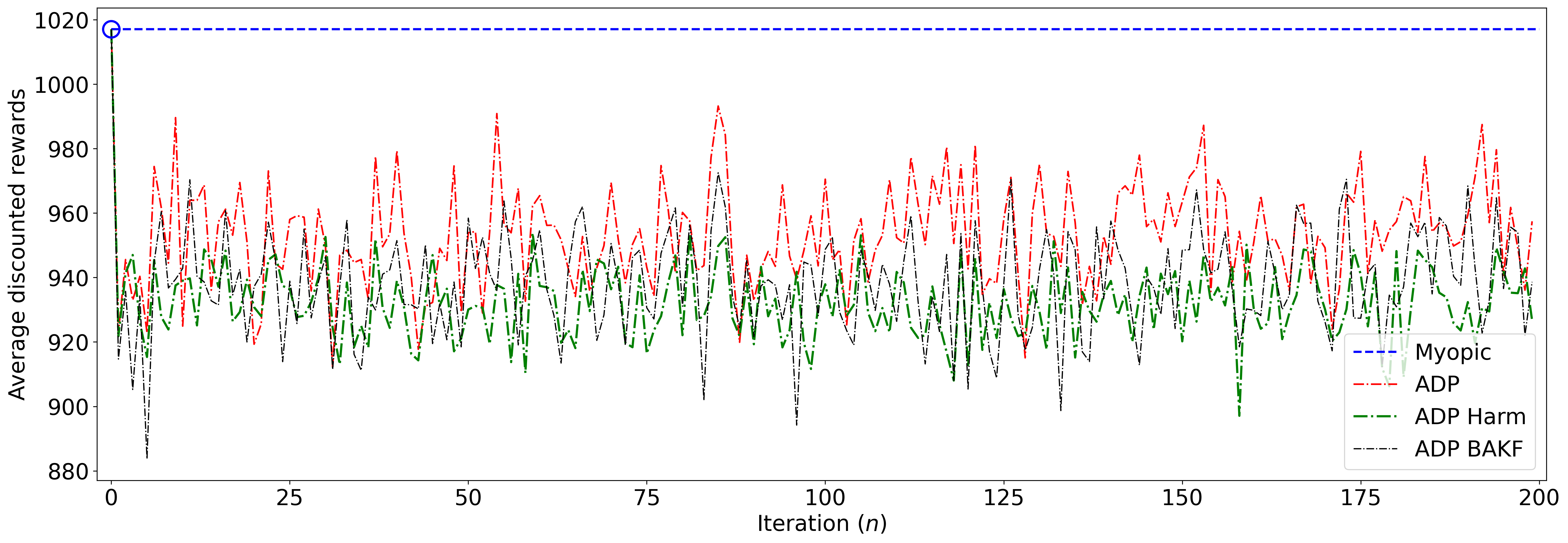}}~
        \subfloat[$\bar{\valuefunction}_0(\state_0)$ Estimate for bug type 1 and due date of -1  - {\scshape{GCC}}\label{fig:estimated_value_post_GCC}]
        {\includegraphics[width=0.5\textwidth]{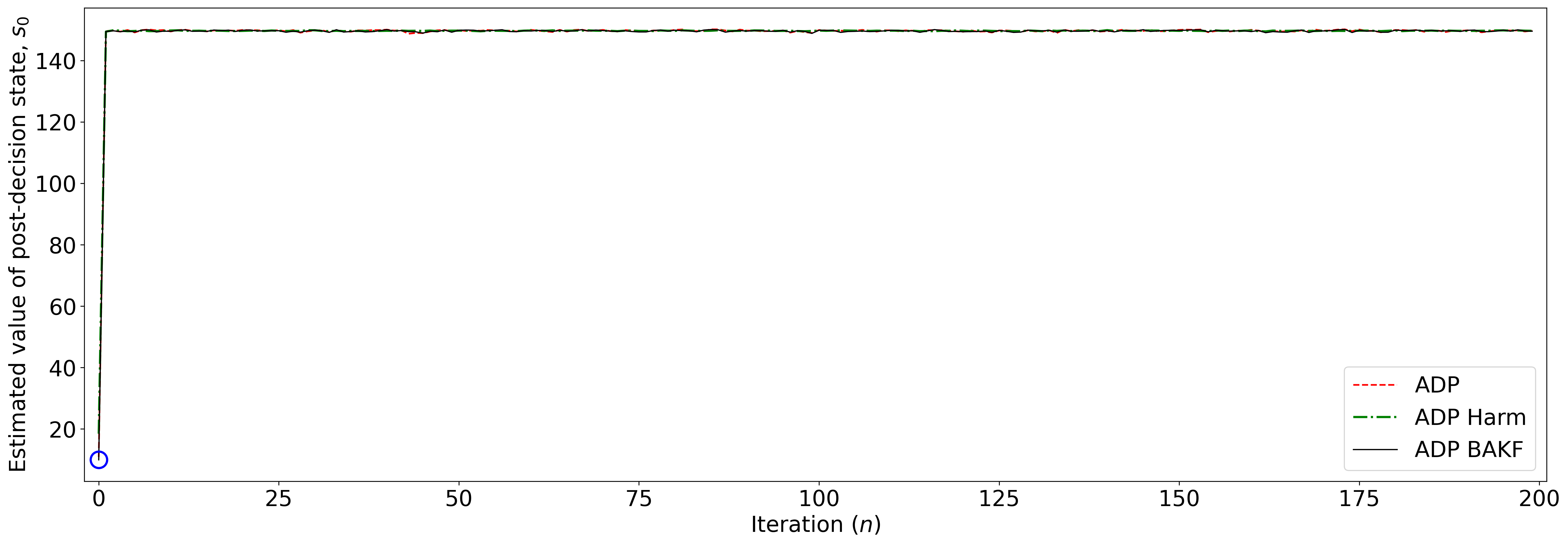}}\\
        \subfloat[Realized rewards - {\scshape{Mozilla}} \label{fig:discounted_reward_Mozilla}]
        {\includegraphics[width=0.5\textwidth]{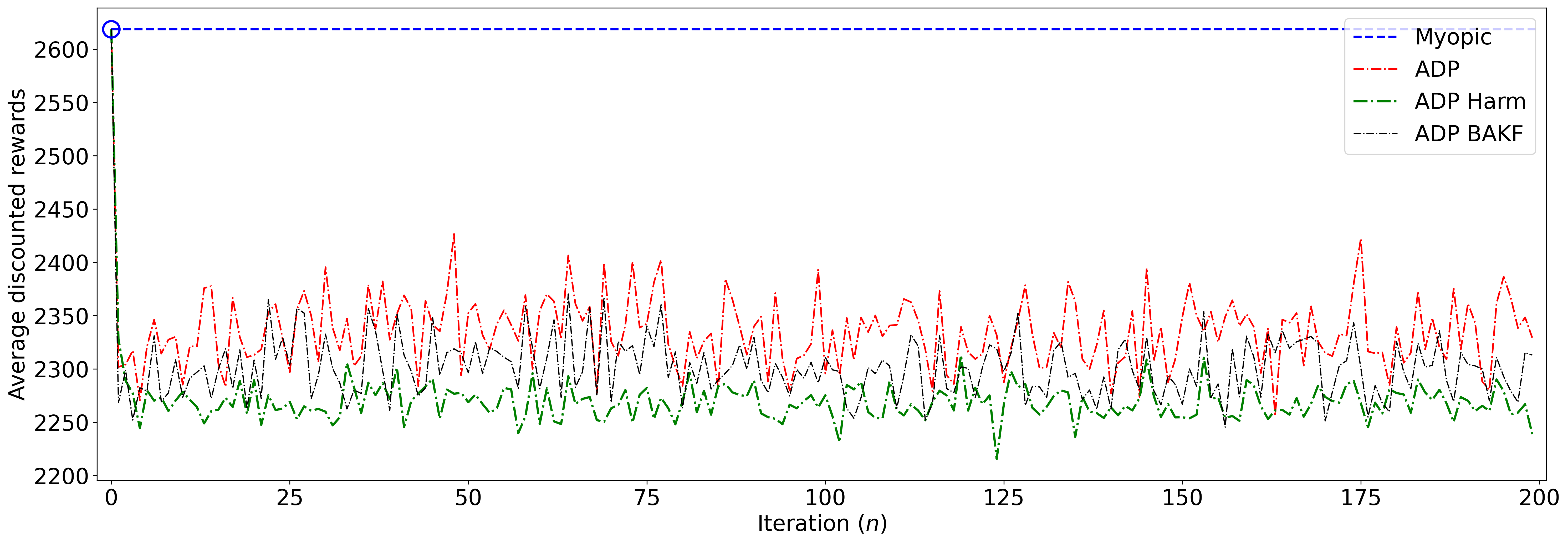}}~
        \subfloat[$\bar{\valuefunction}_0(\state_0)$ Estimate for bug type 1 and due date of -1  - {\scshape{Mozilla}}\label{fig:estimated_value_post_Mozilla}]
        {\includegraphics[width=0.5\textwidth]{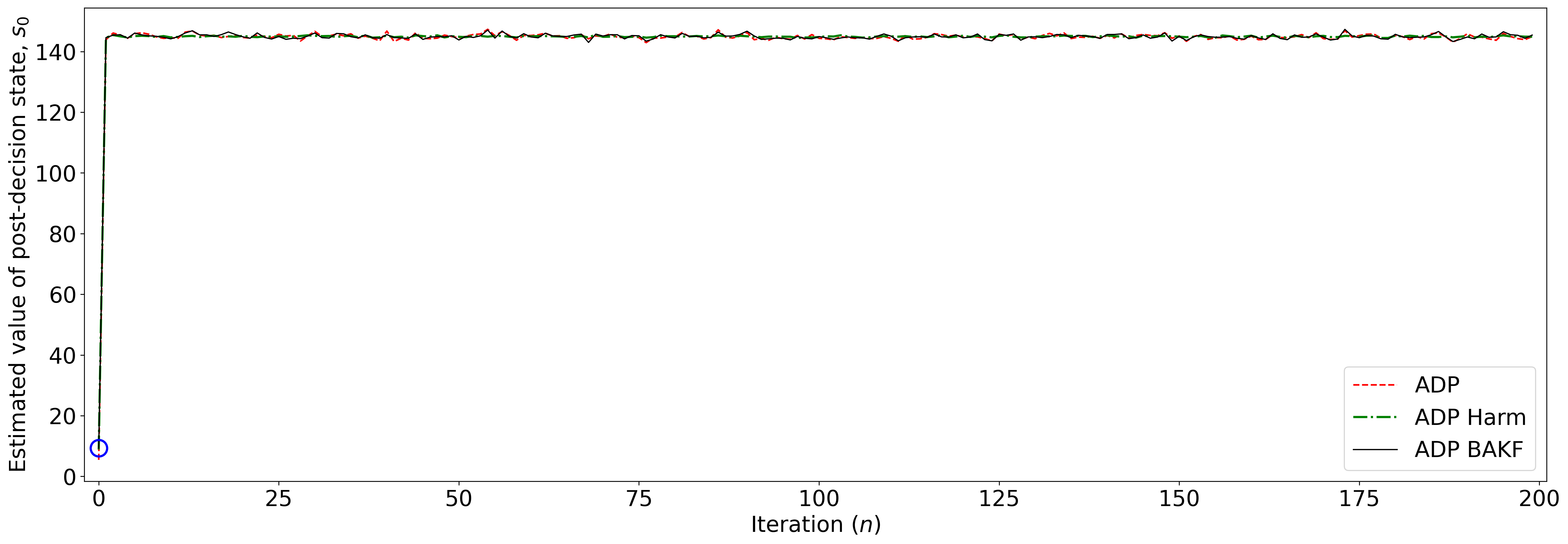}}
        \caption{Finite horizon case: resulting realized rewards (top) and estimated $\bar{\valuefunction}_0(\state_0)$ (bottom) for an arbitrary post-decision state, using $\niteration = 20,000$, $\innerteststep=100$, and $\ninnerepochs=30$.}
        \label{fig:reward_testing}      
  \end{center}
\end{figure*}

We demonstrate two performance measures of the ADP versions. First, we show the estimate $\bar{\valuefunction}^\iteration_0(\StateSpace_0)$ of the initial state $\StateSpace_0$ for different numbers of iterations $\iteration$. Second, for various numbers of iterations $\iteration$, we demonstrate the discounted rewards of employing the learned policy. Specifically, we run a secondary simulation on the side for a certain number of iterations $\iteration$. We ensure that the inner simulation maintains its characteristics for each iteration to make it comparable. Each of these inner simulations has $\ninnerepochs$ epochs, with the value function estimations fixed and the policy based on these values being followed (i.e., we follow the value function approximation from the previous $\iteration$ iterations and do not update the value function estimation during these $\ninnerepochs$ epochs). Accordingly, these inner validation epochs provide insights into the policy enhancements during the training phase. Every $\innerteststep$th iteration, i.e. for $\iteration = 0, \innerteststep, 2\innerteststep, \dots, \niteration$, we run the inner simulation. The Figure~\ref{fig:reward_testing} caption indicates the values used for $\niteration$, $\innerteststep$, and $\ninnerepochs$.

Figure~\ref{fig:reward_testing} show the realized reward of the experiment for different ADP versions. The initial value for the first iteration does not include any presumption about the future costs. Therefore, it is equivalent to the myopic algorithm. As the iterations move forward, the model improves its policy by estimating the future costs of a decision. As a result, we observe a decrease in the average discounted rewards, noting that we have a minimization problem that aims to reduce the bug fixing time. It indicates a better bug assignment policy compared to the myopic version since all the values are smaller than the initial value (the blue line). We observe a higher reduction in discounted rewards for the harmonic and BAKF stepsizes compared to the fixed stepsize ADP. On the other hand, the fixed stepsize demonstrates more fluctuation (i.e., instability) compared to the other two enhancements. We examine the convergence of the value estimation for each ADP algorithm in Figures~\ref{fig:estimated_value_post} and~\ref{fig:estimated_value_post_GCC}. The result indicates a faster convergence of the fixed and BAKF stepsize ADP compared to the harmonic stepsize one in {\scshape{EclipseJDT}}. On the other hand, we do not observe such a behavior for the {\scshape{GCC}} and {\scshape{Mozilla}} projects. Nevertheless, if trained for a sufficient number of iterations, all the algorithms converge to the same value. We report the figure for the post-decision state of an arbitrary bug type and due date. We note that BAKF slightly faster converges to the optimal value and outperforms all other approaches during all iteration steps in the {\scshape{EclipseJDT}} project.

\subsection{Hyperparameter Tuning}
In this section, we investigate the sensitivity of the model to its various hyperparmeters. The most important settings include exploration rate, the cost function for postponement, the discounting factor of the Bellman equation.

\subsubsection{Exploration vs. Exploitation}\label{sec:exploration}

Although previous studies indicate a negligible value in exploration with high-dimensional state space \citep{Kianoush2021, powell2007approximate}, we employ the $\epsilon$-greedy approach to see whether it may help our model to reach a better policy. The $\epsilon$-greedy policy lets the developer decline the assigned by with probability $\epsilon$, or chooses the optimal action $\action$ following the policy $\policy$ with probability 1-$\epsilon$. In the real case scenario, when a developer is CC'ed for a bug, they may start to fix the bug or discard it. This way, we provide the opportunity for the model to visit some states that normally cannot be visited if pure exploitation is taken into account. Table~\ref{tab: exploration} shows how the performance of the models changes when we provide a likelihood of 75 percent for the exploration compared to the case with zero exploration. We observe an improvement in the fixing time and the accuracy of assignments for all methods when $\epsilon$-greedy exploration is employed. We also explored the algorithms during the training phase to see whether exploration helps the models decrease their discounted reward. As discussed in Section~\ref{sec: convergence}, the lower the long-run value for the objective function is, the better the model performance will be. All ADP versions demonstrate a decline/an improvement in their discounted reward when adopting exploration. Accordingly, we use the $\epsilon$-greedy approach with $\epsilon = 0.75$ as the default parameter of our model in the whole paper.

\begin{table*}[!ht]
\centering
\caption{Comparing exploration and pure exploitation\label{tab: exploration}}%
\resizebox{0.75\textwidth}{!}{
    \begin{tabular}{clcrrrrr}
    \toprule
    \textbf{Project} & \textbf{Method} & \textbf{$\epsilon$} & \textbf{Top-1} & \textbf{Top-3} & \textbf{Top-5} & \textbf{Mean Fixing Time} & \textbf{Discounted Reward} \\
    \midrule
    \multirow{6}{*}{\rotatebox[origin=c]{90}{{\scshape{EclipseJDT}}}} & \textbf{ADP} & \multirow{3}{*}{0.75} & 28.2 & 66.7 & 82.5 & 3.93 & 423.43 \\
    & \textbf{ADP Harm} &  & 28.6 & 66.7 & \textbf{82.7} & 3.90 & 417.23\\
    & \textbf{ADP BAKF} &  & \textbf{29.0} & \textbf{66.9} & 82.4 & \textbf{3.88} & \textbf{411.23}\\
    \cline{2-8}
    & \textbf{ADP} & \multirow{3}{*}{0.0} & 28.7 & 65.0 & 80.8 & 3.97 & 423.42\\
    & \textbf{ADP Harm} &  & 27.8 & 65.7 & 81.8 & 3.96 & 420.90 \\
    & \textbf{ADP BAKF} &  & 27.7 & 65.9 & 81.5 & 3.98 & 423.42 \\
    \midrule
    \multirow{6}{*}{\rotatebox[origin=c]{90}{{\scshape{GCC}}}} & \textbf{ADP} & \multirow{3}{*}{0.75} & \textbf{26.1} & \textbf{51.9} & \textbf{59.8} & \textbf{3.59} & 890.00 \\
     & \textbf{ADP Harm} &  & 25.6 & 49.7 & 58.0 & 3.67 & \textbf{851.20} \\
     & \textbf{ADP BAKF} &  & 25.5 & 50.5 & 58.5 & 3.62 & 872.10 \\
    \cline{2-8}
     & \textbf{ADP} & \multirow{3}{*}{0.0} & 25.6 & 48.7 & 56.9 & 3.82 & 957.45 \\
     & \textbf{ADP Harm} &  & 25.4 & 48.5 & 58.1 & 3.76 & 926.66 \\
     & \textbf{ADP BAKF} &  & 25.2 & 48.1 & 56.5 & 3.83 & 938.65\\
    \midrule
    \multirow{6}{*}{\rotatebox[origin=c]{90}{{\scshape{Mozilla}}}} & \textbf{ADP} & \multirow{3}{*}{0.75} & \textbf{6.8} & \textbf{20.1} & 24.5 & 4.85 & 2329.33 \\
     & \textbf{ADP Harm} &  & 6.0 & 19.7 & 24.9 & 4.95 & \textbf{2238.80} \\
     & \textbf{ADP BAKF} &  & 6.2 & 19.7 & 24.2 & \textbf{4.84} & 2313.04 \\
    \cline{2-8}
     & \textbf{ADP} & \multirow{3}{*}{0.0} & 6.7 & 19.9 & \textbf{25.1} & 5.05 & 2281.49 \\
     & \textbf{ADP Harm} &  & 6.3 & 19.2 & 23.1 & 5.48 & 2242.95 \\
     & \textbf{ADP BAKF} &  & 6.4 & 19.3 & 24.0 & 5.31 & 2255.41 \\
     \bottomrule
    \end{tabular}
}
\end{table*}

\subsubsection{The Cost of Postponement}
\begin{figure}[!ht]
\centering
    \includegraphics[width=0.7\linewidth]{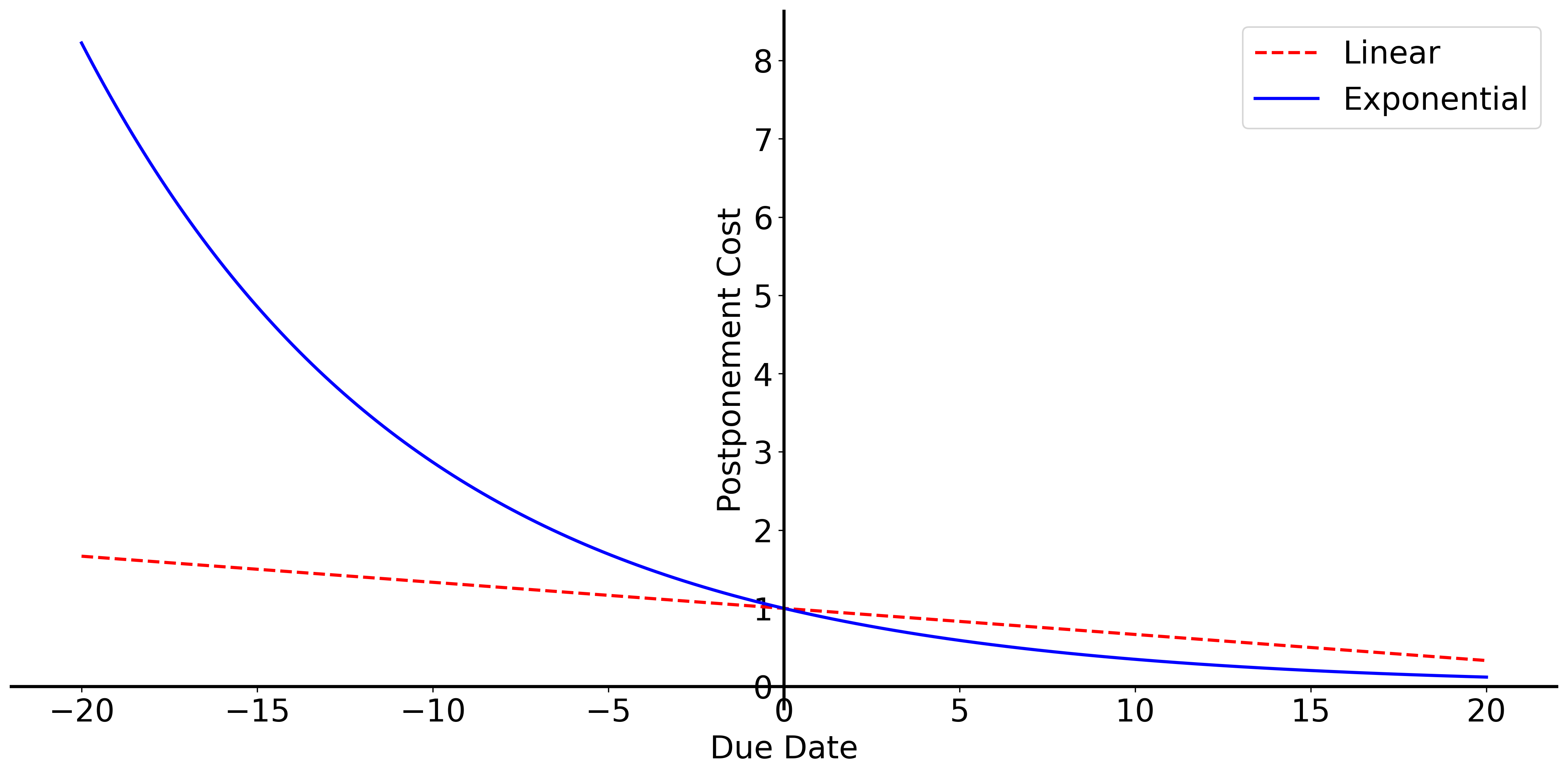}
    \caption{Different function for postponement costs}
    \label{fig:cost_function}
\end{figure}

We define two different cost functions for postponing the bugs (as shown in Figure~\ref{fig:cost_function}), namely Linear and Exponential:
\begin{align}
    \coefficient(\bug_\texttt{due}, \epoch) = 0.9^{\bug_\texttt{due}}. \label{eq: exponential} \\
    \coefficient(\bug_\texttt{due}, \epoch) = \frac{\nepochs - \bug_\texttt{due}}{\nepochs}, \label{eq: linear}
\end{align}
where $\nepochs$ is the project horizon. Using this exponential function (Equation~\eqref{eq: exponential}), we ensure that if we have a large enough number of epochs until the due date of a bug, the postponement cost is low. When we reach the due date, the cost becomes 1, and afterward, it grows exponentially, restricting the model to further delay fixing the bugs whose due dates have passed. Considering the linear function (Equation~\eqref{eq: linear}), the cost would be slightly higher for early assignments and much lower for the late ones compared to the exponential function. We aim to see whether we need to enforce the model by a high cost of delayed assignments or the model learns it through its value function estimation without such a cost.

Table~\ref{tab: posponement} shows the compares linear and exponential cost functions for the bugs postponements. We observe that when we utilize the linear function in the objective function, we have a slight drop in the fixing time and a negligible improvement in the assignment accuracy. Therefore, no matter what cost function is selected, \texttt{ADPTriage} adapts itself to estimate the optimal values of the future rewards. Hence, it is not sensitive to either option. Moreover, there is no difference between the discounted reward of the training phase of the two options. As a result, we use the simpler linear function as the default cost function throughout our experiments.

\begin{table*}[!ht]
\centering
\caption{Comparing different cost functions for postponements\label{tab: posponement}}%
\resizebox{0.8\textwidth}{!}{
    \begin{tabular}{clcrrrrr}
    \toprule
    \textbf{Project} & \textbf{Method} & \textbf{Cost function} & \textbf{Top-1} & \textbf{Top-3} & \textbf{Top-5} & \textbf{Mean Fixing Time} & \textbf{Discounted Reward} \\
    \midrule
    \multirow{6}{*}{\rotatebox[origin=c]{90}{{\scshape{EclipseJDT}}}} & \textbf{ADP} & \multirow{3}{*}{Linear} & 28.2 & 66.7 & 82.5 & 3.93 & 423.43 \\
    & \textbf{ADP Harm} &  & 28.6 & 66.7 & 82.7 & 3.90 & 417.23\\
    & \textbf{ADP BAKF} &  & 29.0 & 66.9 & 82.4 & 3.88 & 411.23\\
    \cline{2-8}
    & \textbf{ADP} & \multirow{3}{*}{Exponential} & 28.1 & 66.2 & 81.8 & 3.96 & 423.43\\
    & \textbf{ADP Harm} &  & 28.7 & 66.3 & 82.1 & 3.92 & 417.23\\
    &\textbf{ADP BAKF} &  & 29.0 & 66.7 & 82.1 & 3.90 & 411.23\\
    \midrule
     \multirow{6}{*}{\rotatebox[origin=c]{90}{{\scshape{GCC}}}} & \textbf{ADP} & \multirow{3}{*}{Linear} & 26.1 & 51.9 & 59.8 & 3.59 & 890.00 \\
     & \textbf{ADP Harm} &  & 25.6 & 49.7 & 58.0 & 3.67 & 851.20 \\
     & \textbf{ADP BAKF} &  & 25.5 & 50.5 & 58.5 & 3.62 & 872.10 \\
     \cline{2-8}
     & \textbf{ADP} & \multirow{3}{*}{Exponential} & 25.7 & 50.6 & 59.0 & 3.63 & 890.00 \\
     & \textbf{ADP Harm} &  & 25.0 & 49.0 & 57.6 & 3.70 & 851.20 \\
     & \textbf{ADP BAKF} &  & 25.2 & 49.9 & 58.2 & 3.65 & 872.10\\
    \midrule
     \multirow{6}{*}{\rotatebox[origin=c]{90}{{\scshape{Mozilla}}}} & \textbf{ADP} & \multirow{3}{*}{Linear} & \textbf{6.8} & \textbf{20.1} & 24.5 & 4.85 & 2329.33 \\
     & \textbf{ADP Harm} &   & 6.0 & 19.7 & \textbf{24.9} & 4.95 & \textbf{2238.80} \\
     & \textbf{ADP BAKF} &   & 6.2 & 19.7 & 24.2 & \textbf{4.84} & 2313.04 \\
     \cline{2-8}
     & \textbf{ADP} & \multirow{3}{*}{Exponential}  & 6.6 & 19.8 & 24.5 & 4.89 & 2329.33 \\
     & \textbf{ADP Harm} &   & 6.1 & 19.7 & 24.8 & 4.97 & \textbf{2238.80} \\
     & \textbf{ADP BAKF} &   & 6.2 & 19.8 & 24.4 & 4.87 & 2313.04 \\
    \bottomrule
    \end{tabular}
}
\end{table*}

\subsubsection{Discount Factor}
\begin{figure*}[!ht]
  \begin{center}
        \subfloat[{\scshape{EclipseJDT}} \label{fig:discounting_EclipseJDT}]
        {\includegraphics[width=0.5\textwidth]{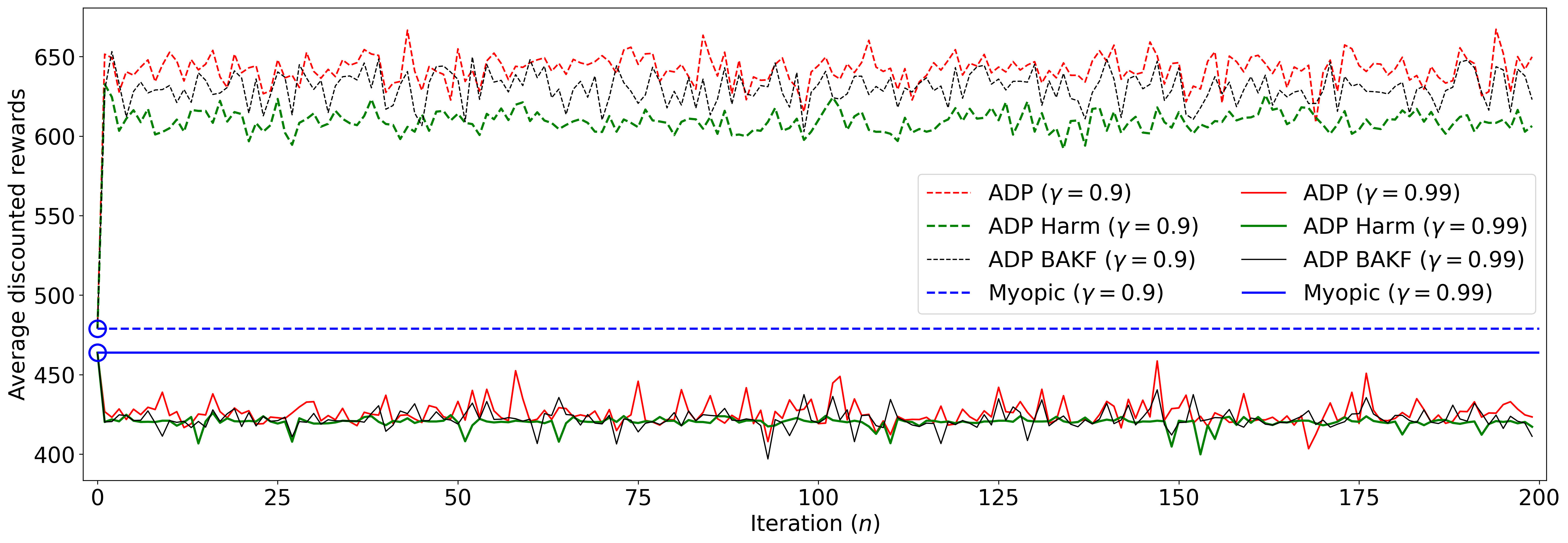}}~
        \subfloat[{\scshape{GCC}} \label{fig:discounting_GCC}]
        {\includegraphics[width=0.5\textwidth]{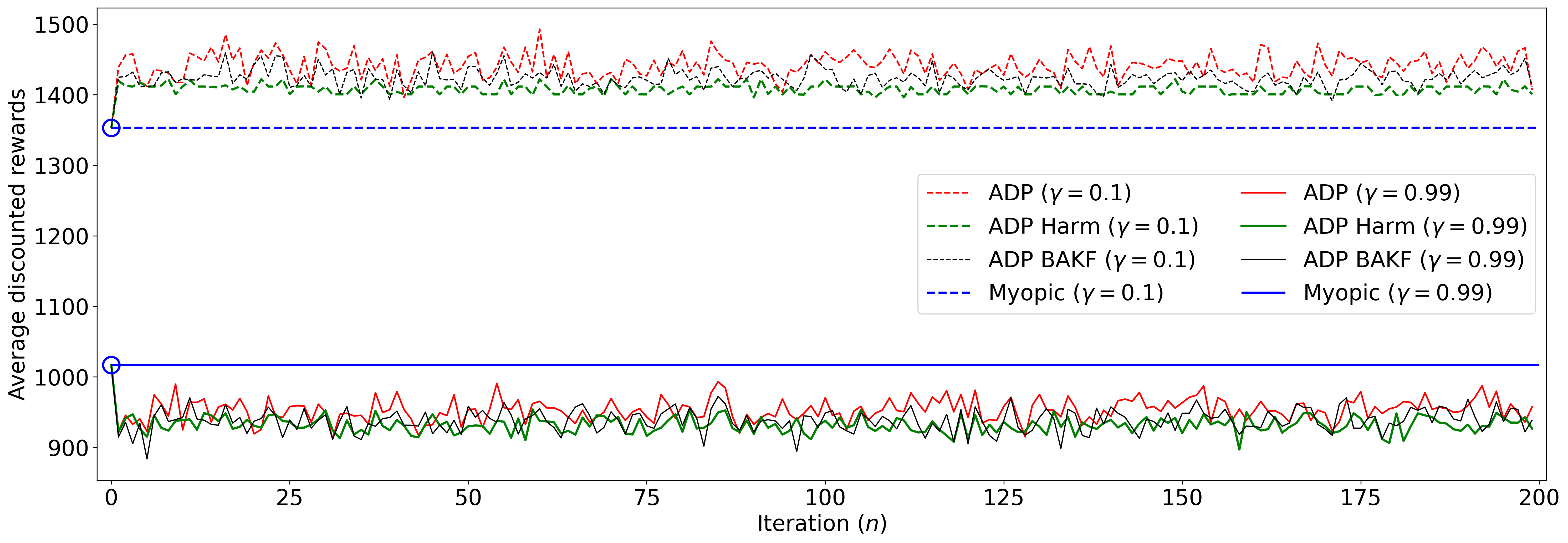}} \\
        \subfloat[{\scshape{Mozilla}} \label{fig:discounting_Mozilla}]
        {\includegraphics[width=0.5\textwidth]{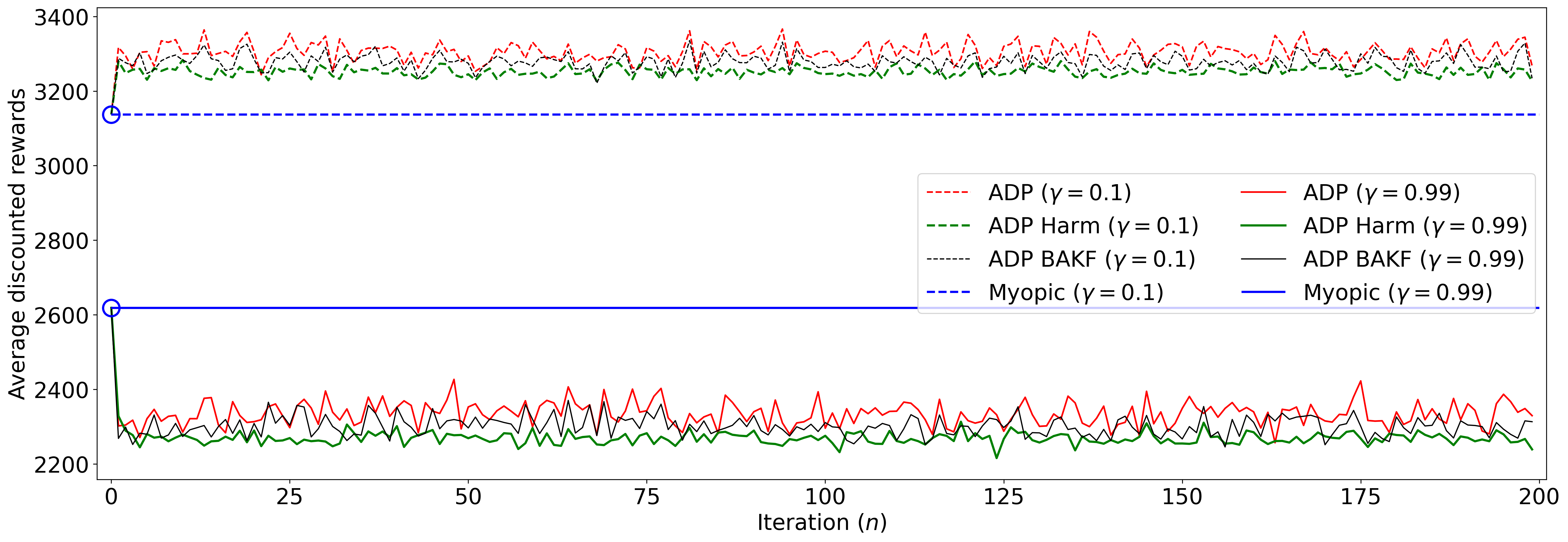}}
        \caption{Sensitivity of the learning to the discount factor. Dashed lines are related to $\gamma=0.9$ and solid lines are associated with $\gamma=0.99$.}
        \label{fig:discounting}      
  \end{center}
\end{figure*}

The discount factor $0<\gamma< 1$ of the Bellman equation controls how we accumulate contributions over time. For $\gamma=0$, we only consider the immediate reward other than the future accumulated rewards. We explore the model sensitivity to two different settings for the discount factor: $\gamma \in [0.9, 0.99]$. Figure~\ref{fig:discounting} shows how the model learns to accumulate the rewards throughout the iterations. We note that for $\gamma = 0.9$, the ADP model underperforms the myopic approach as its discounted rewards are higher than that of the myopic method. As our objective function is a minimization, lower rewards/costs are desired. On the other hand, for $\gamma = 0.99$, after a few iterations, the model starts accumulating lower rewards/costs and learns how to estimate the future costs of a state. Therefore, we utilize $\gamma = 0.99$ based on its promising performance in our numerical experiments. 

\section{Threats to Validity} \label{sec:threats3}
Threats to the validity of our empirical study are as follows.

\subsection{Construct Validity}
We use a train-test split to estimate the models' performance. The first eight years are used as a training period and the latter two years as a test period. However, the results may be influenced by the bug repository's ever-changing nature. Some developers may become less active over some time, leave the system, or become more focused on a single project component. As a result, the active developer's definition may need to be revised yearly. However, we stick to the common practice and definitions to make our results comparable to those of earlier studies. We advocate a rolling strategy for the train-test divide to avoid obsolete conclusions in future research. We note that the difference between the bugs' and developers' frequencies during the training and testing phases has been already reflected in the result.

Although we analyze textual information as our independent feature while finding the LDA category of the bugs, other bug characteristics such as components and keywords can be incorporated into the clustering models. We plan to expand our independent variables and integrate new external impacts in future investigations.

To the best of our knowledge, this is the first research to incorporate uncertainty in bug arrival and developer availability into the modeling process. As a result, comparing the performance of the \texttt{ADPTriage} against that of prior algorithms does not seem feasible. We assume that the developers' availability follows the same distribution as the one in the training and testing phases. We acknowledge the lack of data relating to the developers' real schedule when it comes to the accuracy of constructing the modeling process based on the bugs' history. The same threat exists in estimating the length of time it will take to assign a bug. In practice, that value can be determined according to a bug's perceived priority and severity. While a developer can work on multiple bugs at once, this may lengthen the time it takes for the developer to fix all of them. As a result, because our approach considers bug fixing costs, assignment due dates, and capacities simultaneously, the problem of a developer's availability is alleviated by a longer fixing time. All these assumptions can be replaced with their actual values in practice and do not violate the validity of the model.

\subsection{Internal Validity}
We use the REST API to collect bug information from Bugzilla, including all bug data between January 2010 and December 2019\footnote{\url{https://wiki.mozilla.org/Bugzilla:REST_API}}. The API, however, is restricted to regular users, and access to some bugs is not possible. As a result, we extract all accessible bug reports. We ensure that all publicly accessible bugs are included in our work.

We exclude the severity and priority of bugs in our model because they are usually deemed subjective \citep{akbarinasaji2018partially, Gupta2019, tian2015, Yang2017}. Our model can indirectly comprise these two factors by defining the due date for bug assignments. Accordingly, bugs with higher severity/priority may be given a shorter due date and fixed earlier in the system. On the other hand, in the event of a developer who is not an expert in a certain topic, we anticipate seeing high fixing costs (i.e., low suitability). As a result, the model may defer bugs according to their due dates until a more suitable developer becomes available in the future decision epochs. Another important consideration is the relationship between severity, priority, and resolution time. Previous works also indicate a shorter fixing time for bugs with higher severity and priority~\citep{Sepahvand2020, Zhang2013}. Therefore, our model implies those factors by determining shorter assignment times for more severe bugs. 

\subsection{External Validity}
In this work, we examine well-known projects in Bugzilla. Our conclusion may not be generalizable to all other open software systems, even though the projects are compatible with previous studies. Nevertheless, because the projects chosen are extensive, long-lived systems, our study is less likely to be biased. Replicating this work using different datasets is highly recommended. Finally, we use various performance metrics to account for all models' benefits and drawbacks. 

\section{Related Work} \label{sec:related_works3}

Due to the importance of bug triage in the literature, researchers conducted several related studies during the past decades. They adopted different techniques, e.g., machine learning, graph analysis, fuzzy set-based automatic bug triage, and deep learning techniques. However, the uncertainty in the issue tracking systems is yet to be explored. Specifically, in open-source projects, many free-lancer developers, whose availability in the upcoming period is not easy to determine, may attempt to fix bugs. Moreover, the timing of bug reports may not follow a specific distribution, thus not being straightforward to predict the next bug report~\citep{Jahanshahi2020bugnumber}. 

Previous studies mainly focused on improving the accuracy of the assignment in a static environment. In other words, knowing the bug attributes and previously assigned developers, how we can propose an appropriate developer for a particular bug. For instance, \citet{xuan2017automatic} suggested that using expectation-maximization and a combination of labeled and unlabeled bug reports can improve the performance of the Naive Bayes classifier in the bug assignment task. They used a small sample of labeled bug reports to train a classifier. In another study, \citet{Pedro2021} explored different traditional machine learning techniques using an industrial dataset and reported the accuracy of the assignments for each algorithm. These traditional methods were further improved using deep learning approaches~\citep{mani2019deeptriage, Choquette2019}, whether adopting the same textual information or incorporating team labels or other attributes. These approaches are mainly prone to overspecialization, i.e., assigning numerous bugs to expert developers~\citep{kashiwa2020}. On the other hand, they do not consider the capacity and schedule of each developer while triaging the bugs. The dynamic nature of the ITS and the rate of incoming bug reports require defining an environment in which these features are captured.

\citet{kashiwa2020} first proposed a simulation environment within which they can implement their release-aware bug triage method. That way, they guarantee that their method is applied in a system similar to the actual one. Moreover, \citet{jahanshahi2020Wayback} introduced their tool, called Wayback Machine, to reconstruct the actual bug arrival times and the historical decisions made in the ITS. Although both built a simulated environment to examine their algorithms, none considered the uncertainty in bug arrival times and developers' availabilities. \citet{guo2020developer} emphasized the importance of developers' activities in the bug triage process, i.e., whether a developer remains active in certain bug types during different project phases. They combined these developers' engagements with the textual information of the bugs fed into a Convolutional Neural Network to enhance the bug assignment. However, the rate at which a bug type is reported to the ITS and the developers' schedules are not included in their model. One of the sources of uncertainty, developers' availability, is first considered by ~\citet{jahanshahi2022sdabt}. They proposed an IP solution for bug triage, incorporating bug dependencies, developers' schedules, and bug fixing costs. Nevertheless, the IP model lacks getting updated over time and does not consider the future cost of each bug assignment. These semi-online methods made through simulation are not able to capture the uncertainty in the ITS.

In a more recent work by~\citet{Liu2022}, for the first time, Reinforcement Learning (RL) is leveraged to propose an online solution for the bug triage problems. Based on the textual information of the bug reports, their method assigns them to developers with a likelihood of an ``action'' in RL. Accordingly, the feature information of bug reports is considered the ``state'' in (RL). The probability of a developer being chosen to fix a bug can be determined by analyzing the multidimensional properties; hence, the developer with the highest probability value can be identified. That is, the ``reward'' is defined based on the feedback of the assignments. To achieve the desired result, the model self-trains and adapts to the relationship between bug reports and developers. In their definition of the problem, exogeneous information such as bug arrival or developer availability is disregarded. Therefore, the long-term reward may not represent the actual value of the current assignment. On the other hand, they do not consider the developers' schedules. Overlooking such a constraint may lead to overspecialization. Hence, our model differs from the aforementioned studies in that it utilizes Approximate Dynamic Programming to comprise exogenous information in the ITS, uncertainty in this ever-evolving system, high-dimensional state-action pairs, and constraints on developers' burdens. Our online \texttt{ADPTriage} method contributes to the literature in terms of novel bug triage formulation and improved assignment accuracy.

\section{Concluding Remarks} \label{sec:conclusion3}
In this paper, we develop an online solution to the bug triage using MDPs, which takes into account uncertain bug arrivals and active developers of an open-source ITS. 
We present an ADP policy, called \texttt{ADPTriage}, based on the value function approximation of bugs and developers to incorporate the downstream uncertainty in the bug arrivals and developers' schedule. 
In addition, we develop an optimization-based myopic strategy. This method enables real-time decision-making for bug assignment while taking into account developers' experience, bug type, and fixing time attributes without imposing any constraints on the stochastic process. In addition, we consider various enhancements to the ADP algorithm. To the best of our knowledge, this is the first time an ADP-based solution approach developed for the bug triage problem.

We utilize bug reports of the {\scshape{EclipseJDT}}, {\scshape{GCC}}, and {\scshape{Mozilla}} projects between 2010 and 2020, where the first eight years are taken as the training set, and the remaining ones are as the test set. Using elaborate modeling of the environment, we demonstrate the efficiency and effectiveness of the proposed ADP algorithm and the myopic strategy. 
We conduct a sensitivity analysis on model parameters and find that the ADP policy outperforms the myopic policy for online bug triage under various parameter configurations. 
We also demonstrate that \texttt{ADPTriage} makes intelligent decisions based on the system's characteristics and expected future return.

This study serves as a prelude to various new research avenues in the online bug triage decision-making. 
First, it can be extended to proprietary software systems in which the developers' schedules are more stable, and the chance of improvement over the myopic approach is higher because of such patterns. 
Second, the system state can be expanded to incorporate developers' ability to address simultaneous bug-fixing tasks~\citep{jahanshahi2022sdabt}. 
Since such a modification significantly increases the system state dimensionality, more complex solution algorithms (e.g., Neural ADP) can be employed to overcome that issue~\citep{Shah2020}. 
Lastly, by taking into account the uncertainty in estimated bug fixing times, the enhanced \texttt{ADPTriage} can better reflect the real-world scenarios.  

\section*{Supplementary Materials}~\label{sec:supplementary3}
To make the work reproducible, we publicly share our originally extracted dataset of one-decade bug reports, scripts, and analysis on \href{https://github.com/HadiJahanshahi/ADPTriage}{GitHub}.



\bibliographystyle{elsarticle-harv}
\bibliography{references}

\end{document}